\documentclass[draft2]{aastex62}
\pdfoutput=1 %for arXiv submission
\usepackage{amsmath,amstext}
\usepackage[T1]{fontenc}
\usepackage{apjfonts} 
\usepackage[figure,figure*]{hypcap}
\usepackage{subfigure}
\usepackage{epsfig}
\usepackage{longtable}
\usepackage{isotope}
\usepackage{threeparttable}
\usepackage{nicematrix,tikz}
\usepackage{array}
\usepackage{booktabs}
\usepackage{threeparttable}
\usepackage{tabularx}
\usepackage{multirow}

\usepackage[kerning=true]{microtype}
\SetExtraKerning
    {encoding =  {OT1,T1,T2A,LY1,OT4,QX,T5,TS1,EU1,EU2}} % all text
    {
        \textemdash  = {167,167} % thinspace = 1/6 em
    }

\newcommand{\cutt}[1]{\textcolor{blue}{}}

\newcommand{\Ms}{{\ensuremath{{M}_{\odot} }}}

\shorttitle{Dwarf Galaxies and their CGM}
\shortauthors{Tung and Chen}

\begin{document}

\title{Coevolution of Dwarf Galaxies and Their Circumgalactic Medium Across Cosmic Time}

\author[0009-0000-9401-5470]{Pei-Cheng Tung}
\affiliation{Institute of Astronomy and Astrophysics, Academia Sinica, Taipei 10617, Taiwan} 
\affiliation{Department of Physics, National Taiwan University, Taipei 10617, Taiwan}

\author[0000-0002-4848-5508]{Ke-Jung Chen}
\affiliation{Institute of Astronomy and Astrophysics, Academia Sinica, Taipei 10617, Taiwan} 
\affiliation{Heidelberg Institute for Theoretical Studies, Schloss-Wolfsbrunnenweg 35, 
Heidelberg 69118,
Germany}

\begin{abstract}

Dwarf galaxies are thought of as the building blocks of large galaxies such as our Milky Way. 
This paper presents new high-resolution hydrodynamical simulations of dwarf galaxies and their intergalactic medium with the \texttt{GIZMO} code. Our simulations consider the key physical processes of galaxy evolution, such as gas cooling, chemistry, and stellar and black hole feedback. Unlike the previous work, the initial conditions of our simulations taking the dwarf galaxies of $2-5 \times 10^{10} \, M_\odot$ from the realistic cosmology simulations, \texttt{IllustrisTNG}. We further increase the original resolution of \texttt{IllustrisTNG} by a factor of $\sim 100$ via a particle splitting scheme. Our results show that the evolution of complex multiphase CGM and its metal content is sensitive to the redshift of dwarf galaxies. The accretion of CGM into dwarf galaxies plays a key role in providing $20 \% - 50 \%$ of the star-forming gas and replenishing $40 \% - 70 \%$ of the total mass in the galactic disk. Furthermore, the accretion history of supermassive black holes in the centers of high-$z$ dwarf galaxies shows episodic patterns with high-accreting states close to $\sim 10 \%$ of the Eddington mass accretion rate, implying the rapid growth of supermassive black holes in the early universe, which may be revealed by the coming observations from the James Webb Space Telescope (JWST).

\end{abstract}


\section{Introduction}

Based on modern cosmology, the large scale of the universe grew from the small perturbation seeded by the inflation during the big bang \citep{1980Peebles, 1996Navarro, 2000Cole, 2003Dodelson, 2017Stierwalt}. Due to gravitational instability, dark matter started to cluster and formed into isolated structures known as halos, which are the home of galaxies and galaxy clusters. Dwarf galaxies (DGs) reside in small dark-matter halos and can be treated as the building blocks of larger galaxies. Therefore, studying the evolution of DGs is crucial for understanding the formation and evolution of more massive galaxies, such as the Milky Way. 

The circumgalactic medium (CGM) and the intergalactic medium (IGM) surrounding the DG can affect its evolution significantly. Both models \citep{Stinson2012, Ford2013, Shen2013, Suresh2017} and observations \citep{Bruns2000, Walker2013, Anderson2014} suggest that the CGM contains a highly dynamic multiphase structure of gas density, temperature, and metallicity \citep{2013Tumlinson, Tumlinson2017, 2022Tchernyshyov}. However, the IGM contains hot ionized gas residing in the filament and void of large-scale structures. The IGM comprises most of the baryon and dark matter of the universe and the bedrock of the formation of cosmic structures \citep{2020Macquart}.
Some gas from the CGM and IGM can cool through radiative processes and eventually accrete onto galaxies, fueling star formation.
Meanwhile, feedback from massive stars, their supernovae, and the accreting supermassive black hole (SMBH) returns energy and gas from galaxies to the CGM and IGM.
This in-and-out feedback loop results in a baryonic ecosystem, which can sustain and regulate galactic star formation \citep{2019Voit, 2020Zinger, 2020Terrazas, 2020Oppenheimer, 2020Davies}. Therefore, IGM and CGM can significantly affect the evolution of a galaxy and prevent gas depletion for star formation \citep{Larson1972, Larson1980, Tinsley1980, 2009Prochaska, 2011Faucher, Peroux2020}.

The accretion of CGM and IGM onto a galaxy depends on its halo mass and environment, which are influenced by the galaxy's location and redshift.
\cite{Dekel2003, 2005MNRAS.363} have proposed two distinct accretion modes: cold and hot.
The mode of accretion depends on the competition between the cooling time ($t_\mathrm{c}$) and the free-fall time ($t_\mathrm{ff}$).
If $t_\mathrm{c} < t_\mathrm{ff}$, the accreting gas can cool before being accreted, resulting in cold accretion.
For hot accretion, $t_\mathrm{c} > t_\mathrm{ff}$, the gas remains hot during accretion.
Cold accretion is more efficient because the accreting flow is unshocked with a lower temperature ($T \sim 10^{4} K$).
Since the temperature of the accreting gas is determined by the gravitational potential of a halo, generally speaking, cold accretion occurs in the halos of virial mass, $M_\mathrm{vir} < 3 \times 10^{11} \, M_\odot$ \citep{Dekel2006, Cattaneo2020}, which allows the IGM gas to flow efficiently into the galaxy and form stars.
Therefore, cold accretion is crucial in shaping these DGs of $M_\mathrm{vir} \leq \times 10^{11} \, M_\odot$.

% Revise Note: makedisk and DICE are idealized galaxy model focusing on different scientific goal. 
% \citet{makedisk, DICE} studied the coevolution between DG and CGM by simulating an idealized disk galaxy. 
%\mynote{Idealized disk galaxy models (e.g. \cite{makedisk, DICE}) can be used to study coevolution between DG and CGM}
Previous studies of coevolution between DG and CGM used idealized disk galaxy models (e.g. \cite{makedisk, DICE}).
\cite{zhu2024} examined the effect of ram pressure stripping on an idealized DG through a wind tunnel setup. 
Meanwhile, other related models used cosmological simulations to obtain more realistic initial conditions for DGs and their CGMs. For examples, \cite{2012van} modeled the \textsc{Hi} absorbers to support the evidence of cold accretion and \cite{2020Rey} studied the evolution of ultra-faint field DGs by considering dark-matter-only cosmological simulation. \cite{Zhu_2022} studied the accretion of IGM from cosmic filaments and obtains the galaxy accretion rates as a function of redshift and halo mass. \citet{Tollet_2022} derived the entropy criteria that distinguish accretion in the cold and hot modes based on the NIHAO simulations \citep{2015Wang}. 
Furthermore, the FIRE team \citep{Hopkins_2014, Hopkins_2018, 2023Hopkins} performed a serial study of the galaxy evolution, tracking the baryon evolution of the galaxy \citep{2017Angles} and the CGM based on cosmological zoom-in simulations \citep{2019Hafen, 2020Hafen}. 

Several studies have revealed the influence of CGM/IGM inflow and outflow on the evolution of DGs. \cite{2012Brisbin} shows that most low-mass ($M_\mathrm{vir} \sim 10^{10} M_\odot$) star-forming galaxies are fed by low-metallicity (Z $\sim 0.1 \, Z_\odot$) gas infall, and \cite{2019Hafen} demonstrates that IGM dominates CGM accretion rather than wind-fed from central or satellite galaxies. \citet{2018Christensen} demonstrates that DGs can eject more metal to CGM due to their shallow gravitational potential. Furthermore, about $50 \%$ of the out-flowing gas from DGs falls back onto the host DGs \citep{2016Christensen}. On the other hand, gas outflow from the CGM leaves the host halo and becomes diffuse IGM \citep{2020Hafen}. However, these previous studies did not consider the redshift evolution of DGs and their CGM from z=2 to z=0, which will be soon examined by the JWST observation.

% Revise Note: Move to discussion
% Most CGM gas is ejected into the IGM for low-mass halos around z = 2, with outflow strength strongly dependent on halo mass and redshift \citep{2020Hafen}.

Although these studies provide significant insight into the evolution of DGs and their CGM, some obstacles remain.
The isolated galaxy model is relatively computationally cheaper and thus can reach a higher resolution to resolve the physical processes in the ISM.
However, the assumption of symmetric or isotropic CGM could be oversimplified and thus lead to less realistic results.
However, cosmological simulations can generate galaxies directly from the formation of large-scale structures. Still, their low spatial resolution limits the possibility of accurately resolving the structure of the multiphase CGM, star formation, and associated stellar feedback model, which has been proven to be important in modeling galaxy evolution \citep{Kim2016, 2024Wright}.
% processes.
Cosmological zoom-in simulations like FIRE are feasible approaches to balance between the simulation domain and resolution. 
However, these zoom-in simulations still neglect the cosmic environmental evolution of galaxies, which is critical for the formation and dynamics of the IGM and CGM.

% our study
Therefore, we perform new 3D high-resolution hydrodynamical simulations of DGs using the \texttt{GIZMO} code \citep{Hopkins_2015, Hopkins_2018}. Unlike previous models, we adopt the results from the state-of-the-art cosmology simulation, \texttt{IllustrisTNG} \citep{nelson2021illustristng}, as the initial conditions for our simulations. Our simulations consider the physical processes required for properly modeling the galaxy evolution, such as gas cooling and chemistry, star formation and stellar feedback, and SMBH feedback. In this work, we focus on the coevolution of DGs and their surrounding CGM and IGM at redshifts ($z$) of 0, 1, and 2.
The goal is to better understand the evolution of DGs along with their environment across cosmic time.

We introduce the methodology of our simulations in Section 2, and then present our results in Section 3. In Section 4, we discuss our results and their astrophysical implications. Finally, we conclude our findings in Section 5.

\section{Methodology}

\subsection{\texttt{GIZMO} code}
% A section for GIZMO code and subsections for its micro-physics.
We use the modified version of the hydrodynamical simulation code \texttt{GIZMO}, a descendant of a widely used smoothed-particle hydrodynamics (SPH) code, \texttt{GADGET-2} \citep{Springel2005-gadget2}. 
\texttt{GIZMO} features quasi-Lagrangian numerical schemes with moving-mesh-like calculations for hydrodynamics, including meshless finite volume (MFV) and meshless finite mass (MFM).
The physical quantities of a gas cell are computed by the cubic spline kernel \citep{Monaghan1992} and evolved by solving the flux on the boundaries from the Voronoi tessellation of gas particles.
These schemes combine the advantages of SPH codes, such as the conservation of angular momentum, with the accuracy in flux calculation offered by mesh codes.  
Furthermore, \texttt{GIZMO} also includes comprehensive modules required to model key physical processes in galaxy evolution. 
We describe our simulation setup and associate physics modules in the following sections.

\subsubsection{Hydrodynamics}
% MFM & subgrid mixing
We adopt the MFM method to solve homogeneous Euler equations in the simulations.
This method ensures mass conservation in each gas cell by eliminating the mass flux between gas cells while evolving the fluid.
Based on \cite{Hopkins_2018}, the advantages of MFM are as follows.
First, MFM couples well with the gravity solver, thus reducing the error from the self-gravity calculation of gas particles.
Secondly, MFM is the best algorithm for precisely tracking the gas flow due to the conservation of mass in each gas cell.
Third, MFM can also reduce the numerical errors of the particle-splitting method in our simulations (see Section \ref{ch2:resolution}).

In addition, we also include a subgrid model of small-scale turbulence between gas particles. We adopt an eddy mixing model from \cite{Smagorinsky1963} in our simulations, following \cite{Colbrook_2017, Hopkins_2018, Rennehan_2018}. 
This model evaluates diffusion by calculating the gradient and shear tensor.
This subgrid model can drive the mixing of metallicity, velocity, and energy at particle scale, making the simulation more physical.
\\

\subsubsection{Gas Cooling}
% EOS, cooling, UVB
Our gas cooling includes the combined effect of H and He cooling, collisional processes, free-free emission, molecular hydrogen, dust collisional effects, fine-structure lines, cosmic rays, and Compton effects \citep{Hopkins_2015, Hopkins_2018}. 
Due to the chemical enrichment from supernovae, metal-line cooling is also considered by assuming the solar abundance pattern for the enriched gas. 

The molecular and fine structure cooling is included to allow gas cooling to a temperature of $\sim 10-10^4$ K to better model the galactic ISM and star formation \citep{Hopkins_2018, 2023Hopkins}. 
Instead of solving the chemical evolution, we use a fitting function from \cite{Krumholz2011} to calculate the molecular gas fraction based on the gas density and metallicity. 
The result of the molecular fraction is also applied in the star formation routine (Section \ref{ch2:SF, FB}).
Additionally, the standard UV background is incorporated into the simulations using a UVB table, TREECOOL. The latest version of TREECOOL \citep{Faucher_Gigu_re_2020} includes the redshift correction of UVB and the effects of AGN feedback \citep{O_orbe_2017, Shen_2020}.

\subsubsection{Star Formation and Stellar Feedback}
\label{ch2:SF, FB}

% star formation, stellar feedback
We adopt the direct sink formation of gas cells and the stellar feedback on the surrounding cells as an IMF-averaged star cluster. 
With a decent gas resolution, $\sim 600 \, \Ms$ per cell, we can resolve the relevant physical scale of giant molecular clouds and apply the star cluster formation using the IMF sampling \citep{Revaz2016}.

The star formation routine converts a gas particle into a star particle if the following criteria are met. First, the density of the gas cell must exceed the critical density $n_{\text{crit}} = 1000 \, \text{cm}^{-3}$, optimized for our resolution to prevent unphysical star formation in low-density regions.
Second, the star-forming gas cell must be surrounded by a convergent flow ($\nabla \cdot v < 0$ with the gas velocity, $v$) and maintain self-gravitating \citep{Hopkins_2013}. 
The self-gravitating (gravitational bound) criterion ensures that a gas cell's kinematic and thermal energy is lower than the local gravitational potential.
Eventually, for any gas particle that meets these conditions, the star formation efficiency is scaled by the molecular fraction of gas satisfying self-shielding to undergo star formation \citep{Hopkins_2018}.

After a star particle forms, its stellar feedback influences the surrounding gas. 
We used a stellar feedback model based on the AGORA project \citep{Kim2016}, which accounts for the time- and IMF-averaged stellar feedback from core-collapse supernovae (CC-SNe) in a star cluster over $30$ Myr, as most massive stars have lifetimes shorter than $30$ Myr.
A constant SN event rate is set to determine the number of SNe exploding for a star particle, $N_\text{sn}$ at a given timestep. Then, the total energy of $N_\text{sn} \times 10^{51}$ erg will be dumped onto the gas around the star particle in the form of kinetic energy to prevent overcooling.
The algorithm developed in \cite{Hopkins_2018-fb} adds momentum statistically isotropically with a kernel-weighted evaluation to ensure the conservation of energy and angular momentum.
The momentum and energy distributed to the nearby gas are calculated on the effective faces of the Voronoi cell, constructed between the star particle and the interacting gas cells.
This method can explicitly resolve the structure of the Sedov blast wave, which has been validated in previous studies \citep{Hopkins_2014, Kimm2014, Martizzi2015}. 

\subsubsection{SMBH Growth and Feedback}
% SMBH feedback
Since the spatial resolution of our simulations cannot resolve the accretion disk of SMBH at the center of DG, we use a simplified and spherically symmetric accretion model, Bondi–Hoyle accretion \citep{Bondi1952} for the SMBH accretion. 
The model accretes gas stochastically onto the SMBH with the accretion rate limited by the Eddington rate \citep{Springel2003, Matteo2005}, based on the radiative efficiency $\epsilon_r \approx 0.1$ according to observations.

After the accretion of gas, the kinetic feedback from the disk wind is deposited onto the surrounding gas in the form of momentum \citep{Hopkins2016}. The disk wind of SMBH follows the energy of $0.5\dot{M}_\mathrm{w} V_\text{w}^2$, with a wind velocity of $V_\text{w} \sim 10\%$ speed of light and mass loading of $\dot{M}_\mathrm{w} = 0.5 \, \dot{M}_\mathrm{BH}$, where $\dot{M}_\mathrm{BH}$ is the mass accretion of SMBH. To compensate for the dynamical effect on the SMBH, we also employ a dynamical friction model for the SMBH particle that applies the gravity-dragging force from the medium of gas and dark matter to stabilize the motion of SMBH at the galactic center \citep{Tremmel_2015, Lin2023}.

\subsection{Simulation Setup}

\subsubsection{Initial Conditions from the \texttt{IllustrisTNG} Project}

% The unique advantage to combine TNG ICs with GIZMO simulations?
% sample selection
To obtain the realistic environments of the DG and its environment, we initialize our simulations with the data from the state-of-the-art cosmological simulation, the \texttt{IllustrisTNG} project.
We retrieve a small volume from TNG50-1 \citep{Nelson2019, Pillepich2019}, the highest resolution run, and import it onto our simulations. The volume selection in TNG50-1 obeys the following criteria. 
First, for a given $z$, we select two halos of DGs with virial mass $M_{\text{vir}} \sim 5 \times 10^{10} \, M_\odot$.
These two halos are located in distinct environments within the cosmic web—either at the nexus of filaments or in isolation. 
Since we cover the halos at $z = 0, 1, 2$ of in the TNG50-1 simulation and we extract six halos for our simulations. 
Secondly, we limit our selection to the star-forming and gas-rich halo. 
%of %$M_{\mathrm{gas}} > 2.5 \times 10^{-2} \, M_\text{vir}$ and $M_\star >  2.5 \times 10^{-3} \,  M_\text{vir}$.
Additionally, each of the selected DG halos is required to contain an SMBH at its center.
Third, the host DG halo should comprise $> 85\%$ of the total halo mass to exclude major mergers.
Merger histories from subhalo catalogs are carefully examined to ensure that no major mergers occurred during the evolution of our target galaxies.
To evaluate the environmental and redshift effects of our selected DGs, we show the correlation of stellar half-mass ($M_\star$), gas mass within the half-stellar-mass radius ($M_{\mathrm{gas}}$), and star formation rate (SFR) for our models along with all TNG DGs at a given redshift in Figure \ref{fig:star_gas_tng}. For $z=0,1$ DGs, the values of $M_{\mathrm{gas}}$, $M_\star$, and SFR of our models are close to the median values. For $z=2$ DGs, our models deviate more from the median values but are still within one sigma. This also implies DGs of the same stellar mass at high-z can have a larger scatter in $M_{\mathrm{gas}}$ and SFR based on our selection criteria. If the gas-to-stellar mass ratio ($M_{\mathrm{gas}}/M_\star$) of a galaxy is directly correlated with its environment, then for each redshift, our two models, \textbf{a} and \textbf{b}, are separated by the median value of $M_{\mathrm{gas}}/M_\star$, representing galaxies in relative gas-rich and gas-poor environments, respectively.

\begin{figure}
    \centering
    \includegraphics[width=1\textwidth]{gas_star_TNG_new.png} 
    \caption{
    The correlation of $M_\star$, $M_{\mathrm{gas}}$, and SFR for DGs with $M_\star = 10^7-10^9 \Ms$ in TNG50-1. The panels from left to right show DGs at $z = 0$, $1$, and $2$, respectively. Our models are annotated with circles and triangles with color-coded SFR in the plots. Most of our models lie within one standard deviation (dotted lines) from the solid lines representing the a median values of $M_{\mathrm{gas}}$ for a given $M_\star$. Two DGs at $z=2$ are slightly away from the solid line but remain close to one $\sigma$.}
    \label{fig:star_gas_tng}
\end{figure}

After deciding on the targeted halos in \texttt{IllustrisTNG}, we retrieve a spherical volume centered on the selected halo and import it into \textsc{GIZMO} as initial conditions. The radius of the sphere is 440 ckpc, corresponding to at least $\sim 3\,R_\text{vir}$. 
This radius is larger than the standard box size for cosmological zoom-in simulations suggested by \cite{Onorbe2013}.
Finally, we evolve our simulation for $1.5$ Gyr, which is more than $10$ dynamical time, to guarantee the dynamical equilibrium in the systems.
We summarize our initial conditions in Table \ref{tb:ics}.
% \mynote{Since $M_\star$ in Figure \ref{fig:star_gas_tng} are stellar half-mass, it should be different from the $M_\star$ on the Table.}

\subsubsection{Resolution}
\label{ch2:resolution}
% Splitting particles for higher resolutions
To further increase the mass resolution of the initial conditions from TNG50-1, we apply the super-Lagrangian refinement introduced in \cite{Hopkins_2015}, an algorithm for the splitting and merging of particles during the simulation, which shares a similar idea as the adaptive mesh refinement (AMR) for mesh codes. 
A similar method is also used in the project GIBLE \citep{2024Ramesh_1, 2024Ramesh_2} to study the evolution of Milky Way-like galaxies with the \texttt{AREPO} code \citep{Springel2010, Weinberger2020} based on the initial conditions from TNG50-2. 
This method can increase the resolution in both dark matter and gas particles, and it works accurately with the MFM method. 
To ensure better spatial and mass resolution for resolving the ISM physics within a galaxy, we further implement a static and radial-dependent hierarchy refinement in our simulations. The particles within 50 ckpc have the highest resolution, and then the resolution decreases by a factor of 2 for every 50 ckpc starting from the center of the halo defined by the SMBH. 

In our simulations, we set the total particle number of baryons and dark matter to a similar number.
In the central region, the optimal mass resolution reaches $600 \, M_\odot$ for baryon particles and $10^4 \, M_\odot$ for dark-matter particles.
The baryon mass resolution in our study is carefully selected to resolve the galactic ISM and star formation. 
Based on the smoothing length of a particle, the highest spatial resolution is $\sim 4$ pc for gas and $\sim 140$ pc for dark matter.
For the star-forming region, the spatial resolution of the gas is $\sim 5$ pc on average, which ensures the convergence of the star formation.
\\
\\

\begin{deluxetable*}{l|ccccccc}
\tablecaption{Summary of initial conditions \label{tb:ics}}
\tablewidth{0pt}
\tablehead
{
\colhead{Model} & \colhead{$z$} & \colhead{$M_\text{vir}$} & \colhead{$R_\text{vir}$\tablenotemark{$\dagger$}} & \colhead{$M_\star$}  & \colhead{$R_{\star,1/2}$} & \colhead{$M_\text{BH}$} & \colhead{Characters of the dwarf galaxy and its environment} \\
\colhead{} & & \colhead{[$10^{10} \, M_\odot$]} & \colhead{[kpc]} & \colhead{[$10^8 \, M_\odot$]} & \colhead{[kpc]} & \colhead{[$10^6 \, M_\odot$]} & Shape and surrounding CGM/IGM
}
\startdata
\textbf{z0a} & $0$ & $2.84$ & $64.34$ & $5.22$ & $3.26$ & $3.29$ & \multicolumn{1}{p{7.5cm}}{This galaxy has spiral arms and locates \linebreak in relative isolated with hot CGM \& IGM.} \\
\textbf{z0b} & $0$ & $5.09$ & $78.14$ & $4.14$ & $0.89$ & $5.32$ & \multicolumn{1}{p{7.5cm}}{Structure of this galaxy looks irregular and it is surrounded by filaments of diffuse gas.} \\
\hline
\textbf{z1a} & $1$ & $3.97$ & $49.02$ & $1.35$ & $1.62$ & $1.74$ & \multicolumn{1}{p{7.5cm}}{This galaxy looks irregular and it moves away from filaments of diffuse gas.}\\
\textbf{z1b} & $1$ & $4.55$ & $51.30$ & $2.54$ & $1.18$ & $1.58$ & \multicolumn{1}{p{7.5cm}}{The structure of galaxy looks irregular and it is close to a nearby massive galaxy located at $\sim200$ kpc away.} \\
\hline
\textbf{z2a} & $2$ & $6.30$\tablenotemark{$\ddagger$} & $40.29$ & $3.32$ & $1.61$ & $1.88$ & \multicolumn{1}{p{7.5cm}}{The structure of galaxy appears irregular and it is surrounded by gas rich filaments undergoing a minor merger with a smaller halo.}  \\
\textbf{z2b} & $2$ & $9.63$ & $46.42$ & $3.66$ & $1.81$ & $1.81$ & \multicolumn{1}{p{7.5cm}}{The structure of galaxy appears irregular and it is located in nexus of several filaments.} \\
\enddata
\tablecomments{Summary of all halo properties, including model name, redshift ($z$), virial mass ($M_\text{vir}$), virial radius ($R_\text{vir}$), stellar half-mass ($M_\star$), stellar half-mass radius ($R_{\star,1/2}$), and mass of the central SMBH ($M_\text{BH}$).}
\tablenotetext{\dagger}{The virial radius is defined by $R_{200}$, which matter density reaches about 200 times of the critical density of matter in the universe.}
\tablenotetext{\ddagger}{For a DG at $z = 2$, its halo is not fully virialized. Therefore, the virial mass calculated directly by the enclosed mass of $R_\mathrm{vir}$ would overestimate the actual mass. The gravitationally bounded mass for \textbf{z2a} and \textbf{z2b} are $5.13 \times 10^{10} \, M_\odot$ and $9.06 \times 10^{10} \, M_\odot$, respectively.}
\end{deluxetable*}

\section{Coevolution of Dwarf Galaxies and Their CGM}

% 1. physical properties of DGs and CGM: Density projection plot
\subsection{Physical Properties of Dwarf Galaxies and Their CGM}

\begin{figure}
    \centering
    \includegraphics[width=0.5\textwidth]{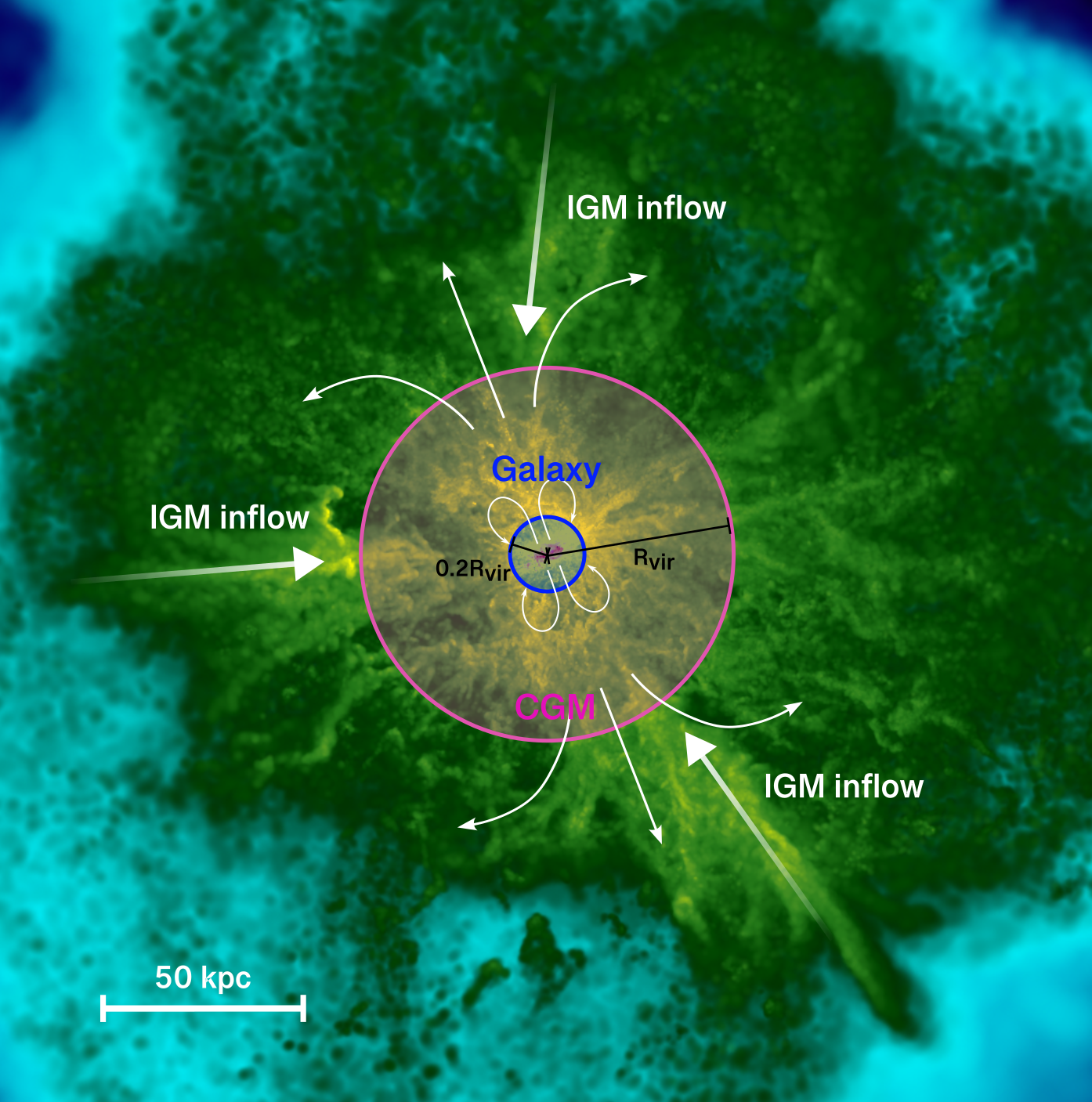} 
    \caption{The schematic figure illustrates the region of the galaxy, CGM, and IGM. The most inner region of $0-0.2 \, R_{\text{vir}}$ (blue), intermediate ring of $0.2 \text{--} 1 \, R_{\text{vir}}$ (pink
), and the outer region of $> \, R_{\text{vir}}$ correspond to the galaxy, CGM, and IGM, respectively. The baryon cycle of the galaxy is demonstrated by the white arrows, indicating the inflow from the IGM, the outflow from the halo, and the recycling of gas inside the CGM.}
    \label{fig:schematic}
\end{figure}

\begin{figure}
    \centering
    \includegraphics[width=1\textwidth]{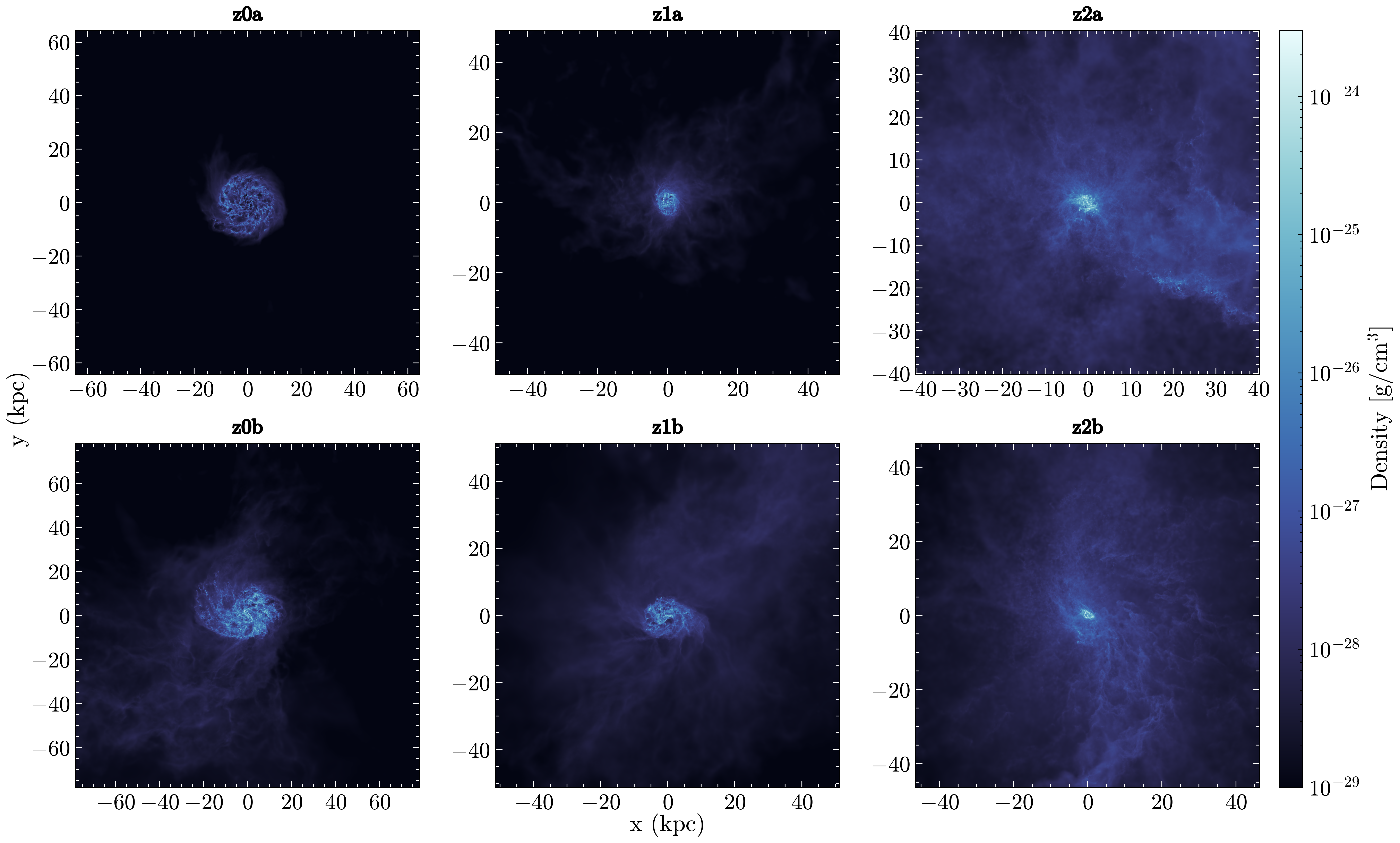}
    \caption{Gas density inside $R_\mathrm{vir}$ for all models at the end of the simulations. All galaxies show some diffuse structures around the major disk. \textbf{z2a} and \textbf{z2b} models show the most prominent diffuse structures due to the strong accretion environment. On the contrary,  \textbf{z0a} residing in gas gas-poor CGM/IGM environment shows little diffuse structures.}
    \label{fig:density}
\end{figure}

Before we present the physical properties of the DGs and their CGM, we first define the region of the galaxy, CGM, and IGM based on their location relative to the halo center, as shown in Figure \ref{fig:schematic}. During the simulation, the inflow and outflow occurring in the galaxy can change the original distribution of gas in galaxy, CGM, and IGM. 
Figure \ref{fig:density} shows the density snapshot at the end of our simulations. Our DGs show irregular structures because of their evolution via accretion flows from the CGM. The redshift of DG has a strong influence on its evolution. 
Based on the gas density distribution, the sizes of the gaseous disks increase from $z = 2$ to $z = 0$. So, the structure of the galaxy is more compact at $z = 1 $ and $z = 2$, while the galaxy at $z = 0$ is relatively sparse and exhibits spiral arm-like structures, particularly for the DG of \textbf{z0a}, which is initially located in a more isolated environment with little surrounding gas.
The CGM of \textbf{z2a} and \textbf{z2b} are filled with dense gas filaments, driving a strong gas accretion onto galaxies and shaping their irregular structure.

% About phase diagram

\begin{figure}
    \centering
    \begin{tabular}{ccc}
    %\centering
        \includegraphics[width=1\textwidth]{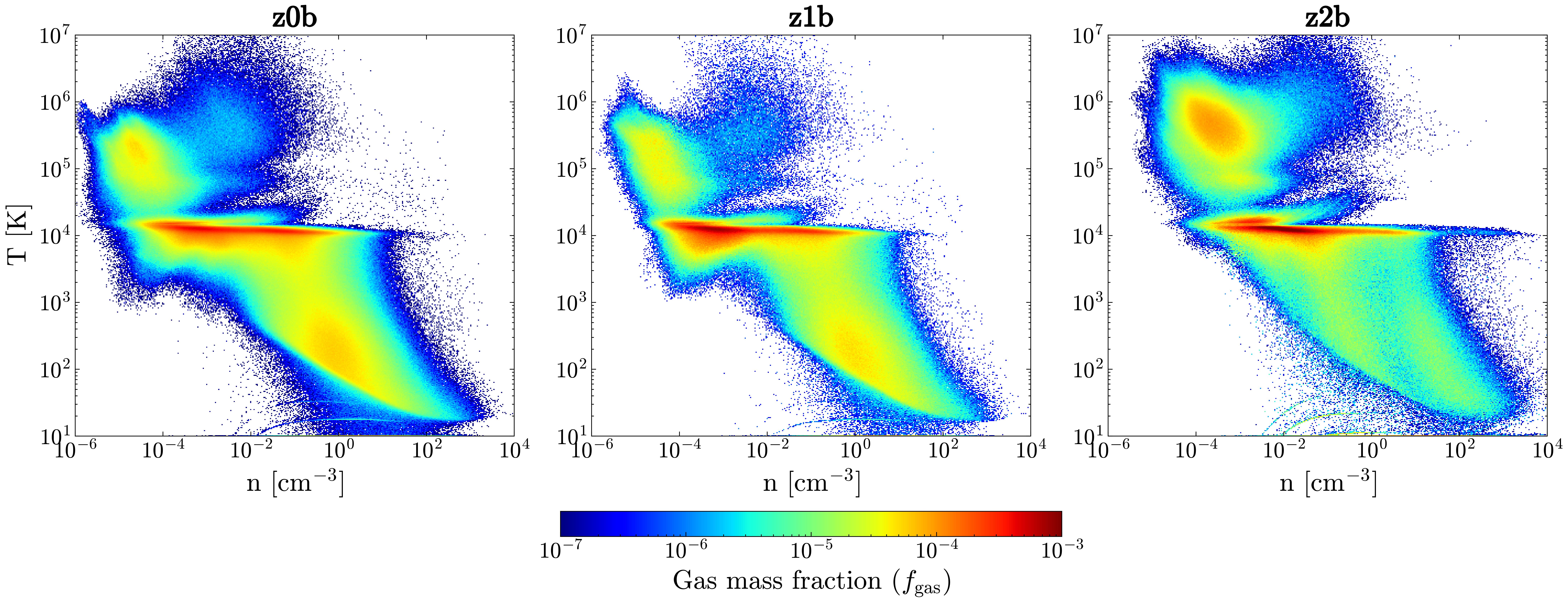} 
    \end{tabular}
    \caption{The temperature-density phase diagram of gas within $R_\mathrm{vir}$ at the end of the simulations. Despite the differences in redshift and environment, the distributions of gas density and temperature show similar patterns among all simulations. Based on the temperature, the gas can be divided into three phases: the cold phase (T $< 10^3$ K), the warm phase (T $\sim 10^3-10^4$ K), and the hot phase (T $> 3 \times 10^4$ K). 
    In each phase, the density and temperature shows a log-normal distribution.
    Most of the gas in both the galaxy and the CGM resides in the warm phase, exhibiting a wide range of density distribution.  }
    \label{fig:phase}
\end{figure}

% gas phase
To explore the physical properties of the galaxy and CGM, we present the gas temperature-density phase diagram for \textbf{z0b}, \textbf{z1b}, and \textbf{z2b} models in Figure \ref{fig:phase}. 
The distribution of gas temperatures and densities roughly follows the trend of pressure equilibrium, extending from top-left to bottom-right through the phase diagram, particularly for the CGM gas of $n \lesssim 10^{-3} \, \mathrm{cm}^{-3}$, as well as for the warm and cold galactic gas with temperatures $< 10^4$ K. 
Furthermore, with temperatures slightly above $10^4$ K%$10^5$ K
, the shock-heated gas follows an adiabatic process, extending from the top right to the bottom left, which is more pronounced in \textbf{z2b} because of a denser CSM at high $z$. 
However, these phase diagrams show noticeable differences among $z$. The CGM of \textbf{z1b} contains the largest amount of cool gas of $n < 0.01 \, \mathrm{cm}^{-3}$ among all models. 
Meanwhile, the CGM of \textbf{ z2b} mostly contains warm or hot gas as a result of heating of the active star formation and the AGN activity.  Additionally, the distribution of cold gas bifurcates in density by showing two yellow stripes for gas of $T< 10^4$ K, which may be caused by the clumpy gas in the disk and CGM.

\begin{figure}
    \centering
    \begin{tabular}{ccc}
    %\centering
        \includegraphics[width=1\textwidth]{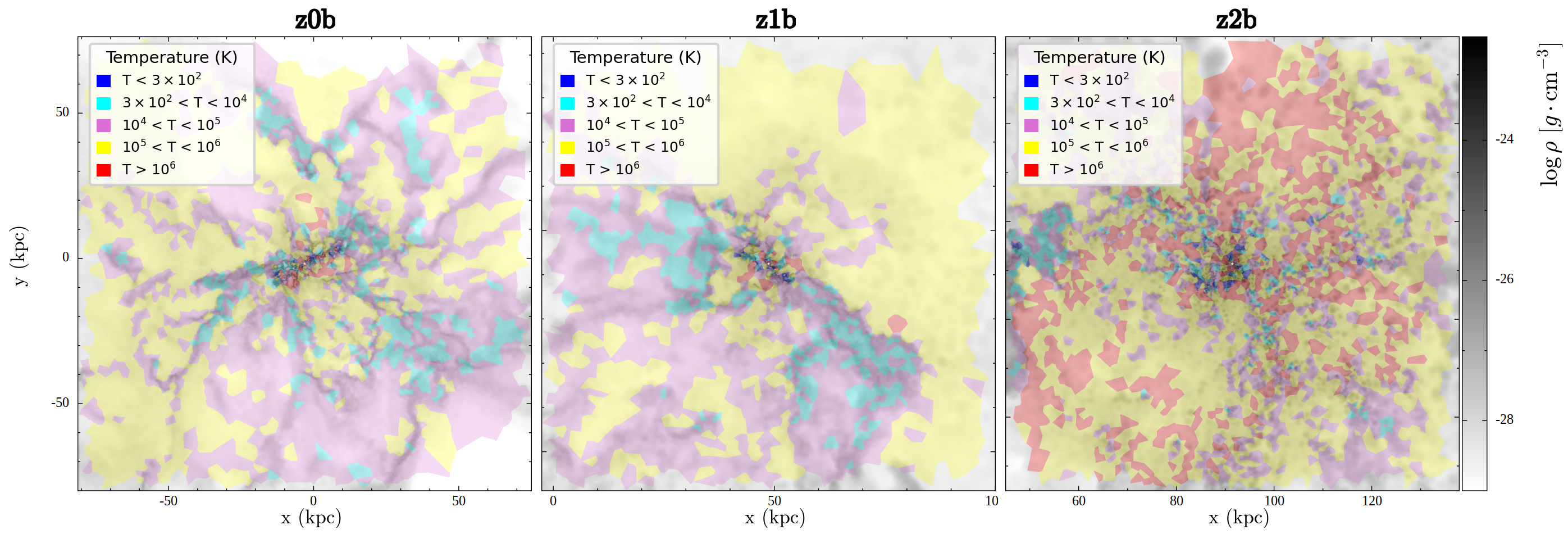} 
    \end{tabular}
    \caption{The temperature distribution of halo gas at the end of the simulations on top of the gas density with grey color.  We show the gas into five different temperature ranges: $\text{T} < 300 \, \text{K}$ (blue), $300 \, \text{K} < \text{T} < 10^4 \, \text{K}$ (cyan), $10^4 \, \text{K} < \text{T} < 10^5 \, \text{K}$ (purple), $10^5 \, \text{K} < \text{T} < 10^6 \, \text{K}$ (yellow), and $\text{T} > 10^6 \, \text{K}$ (red). The patchy color distribution of temperatures and density clumps suggest the highly complex multiphase structure of CGM.}
    \label{fig:phase-temp}
\end{figure}

To investigate the multiphase structure of the CGM, we plot the halo gas temperatures in Figure \ref{fig:phase-temp}.
Most of the gas in the galactic disk has a temperature of $T \lesssim 10^4$ K, and some cold gas can extend into the CGM.  
The distribution of gas phases also depends on the redshift. For the CGM of galaxies at $z = 0$ and $z = 1$, most of the gas is at $T < 10^5$ K, with some filamentary structures of $T\sim10^4$ K connected to central galaxies. This picture is consistent with the cold accretion model \citep{Dekel2003}.  
However, the CGM at $z=2$ is dominated by hot gas. Inside the hot gas, a small fraction of cold gas forms a ring-shaped structure surrounded by the galaxy, which implies the tidal stripping of cool ISM driven by the CGM accretion shown in Figure \ref{fig:phase}.  

We further estimate the $t_\mathrm{ff}$ and $t_\mathrm{c}$ of our models to decide whether the accretion is in hot or cold mode. $t_\mathrm{ff}$ and $t_\mathrm{c}$ are estimated by 
\[t_\mathrm{ff} = \sqrt{\frac{3 \pi}{32 G \rho}}\]
\[t_\mathrm{c} = \frac{\frac{3}{2}kT\cdot n}{n^2 \Lambda}\]
where $\rho$ and T are the averaged temperature and density of the CGM.
The average gas cooling rates is assumed to be $\Lambda \sim 10^{-22} \ \ \mathrm{erg} \, \mathrm{cm}^{3} \, \mathrm{s}^{-1}$ based on the cooling curves from \cite{2007Maio}. With the temperatures and densities from our models, $t_\mathrm{ff}$ roughly ranges from 500 to 1500 Myr and decreases with redshift, while the $t_\mathrm{c}$ ranges from 50 to 1000 Myr, except for \textbf{z0a} where $t_\mathrm{c}$ approaches 1500 Myr due to low-density CGM. For each model, $t_\mathrm{c} <t_\mathrm{ff}$ suggests that cold accretion dominates in these DGs. 
In addition, the ratios of cold to hot gas (separated by $3 \times 10^4$ K) inside the CGMs are $1.04$ (\textbf{z0a}), $3.31$ (\textbf{z0b}), $0.98$ (\textbf{z1a}), $4.3$ (\textbf{z1b}), $5.78$ (\textbf{z2a}), and $1.84$(\textbf{z2b}). These results further support that cold gas dominates the gas inflow, which is in agreement with theoretical predictions from previous studies. The CGM at $z = 2$ remains cold-gas-dominated in mass, despite a large spatial covering fraction of hot gas. However, the fraction of accreted cold gas still varies depending on the local environment. For instance, the isolated environment of \textbf{z0a} and tidal stripping by a nearby massive galaxy in \textbf{z1a} can reduce cold gas accretion, leading to a lower cold gas fraction.

% 2. Gas Accrections of CGM: Accretion rate at 0.2, 1 virial radius
\subsection{Gas Accretion of IGM and CGM}
\label{subsec:gas accretion}

\begin{figure}
    \centering
    \includegraphics[width=1\textwidth]{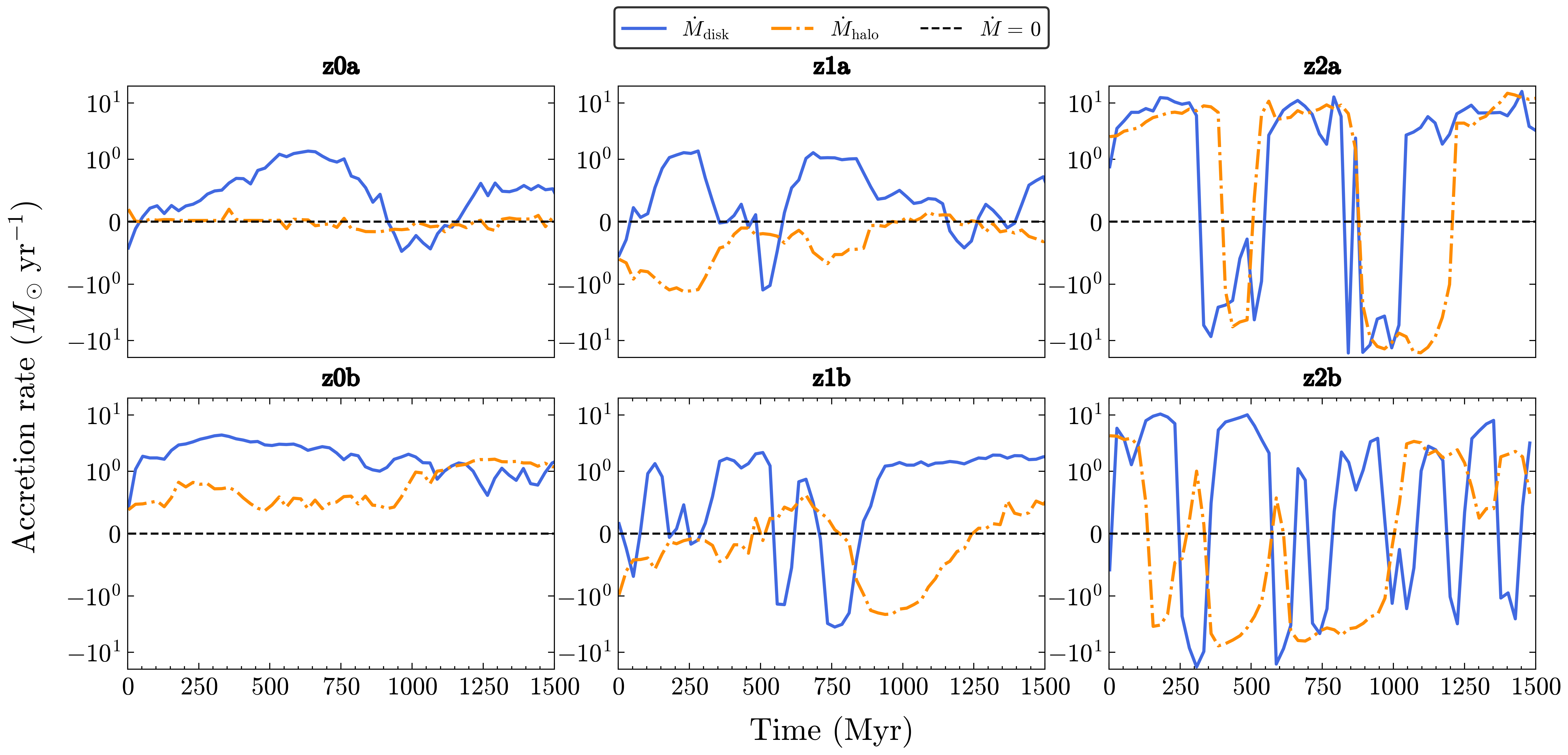} 
    \caption{The evolution of mass accretion rates during the simulation. We calculate the mass accretion rate by temporal variation of the enclosed mass of galactic disk scale within $< 0.2 \, R_\mathrm{vir}$ (blue solid line) and of a halo scale of $R_\mathrm{vir}$ (orange dash-dotted line) along with the reference of $\dot{M}=0$ (black dashed line) . This mass accretion includes both mass inflow and outflow of the targeted scale. Therefore, the positive accretion rate means the inflow mass is larger than the outflow, and the negative accretion rate shows the inflow mass is smaller than the outflow. For all models, the accretion rate at a galactic disk scale of $0.2 \, R_\mathrm{vir}$ is overall positive, indicating the growth of the galaxy with continuous gas fueling. In contrast, the mass accretion rate at $R_\mathrm{vir}$ can be negative due to strong outflows. The accretion of $z = 1$ and $z = 2$ models demonstrate significant variation in both galactic disk-scale and halo-scale.}
    \label{fig:accretion}
\end{figure}

The coevolution of the galaxy and its CGM is critically determined by the accretion of CGM and the outflow from the galaxy driven by the stellar feedback and AGN activity. 
We present the gas accretion and outflow history for the galaxy and CGM in Figure \ref{fig:accretion}. 
The evolution of CGM accretion depends on the redshifts. At $z=0$, the accretion rate is relatively steady. The weak outflow from halo in \textbf{z0a} results from the limited gas supply from an isolated environment. On the other hand, \textbf{z0b} maintains a stable accretion rate from the cold accretion of filaments.  
For galaxies, their gas accretion rate varies significantly with time, especially at the galactic scale during the first 900 Myr. For the scale of $\geq R_\mathrm{vir}$, the gas outflow exceeds the gas accretion most of the time. Furthermore, we find that the accretion rates of the galactic scale region and the halo scale region show a trend of anti-correlation for \textbf{z1a}, \textbf{z1b}, and \textbf{z2b} in Figure \ref{fig:accretion}.
The galactic disks in \textbf{z2a} and \textbf{z2b} show the strongest gas accretion rate, with a peak value around $10$ times higher than those at $z = 0$ and $z = 1$. 
Again, accretion rates vary significantly with time.  
The accretion and outflow rates fluctuate between 10 and -10 $M_\odot \, \mathrm{yr}^{-1}$. Unlike galaxies at $z = 1$, the accretion rate at galactic and halo scales is correlated but with a time offset.  

The differences in accretion rates result from the combined effects of redshifts, environments, and feedback as shown in Figure \ref{fig:accretion}. For DGs at higher redshifts, their surrounding CGM and IGM densities are higher, naturally leading to higher accretion rates. DG halos at $z = 2$ achieve the highest accretion rates. In contrast, at $z = 1$, galaxy clustering substantially affects our target DG halos at the CGM scale, suppressing cold gas inflow from both the CGM and the IGM.
%The environment surrounding these halos also significantly influences their accretion rates, varying with redshift. 
%Embedded within dense cosmic filaments and surrounded by numerous smaller halos, 
Additionally, lower-mass galaxies exhibit more bursty star formation compared to higher-mass galaxies \citep{2012Weisz}. Our $z = 1$ models represent an evolution path transiting between the high accretion rates as $z = 2$ to $z = 0$.  During this regime, galactic-scale gas inflow influences halo-scale accretion after a delay due to episodic star formation and subsequent stellar feedback. Such delayed interactions create an anti-correlation between accretion rates at halo and galactic scales. A similar phenomenon is also observed during the early evolution of \textbf{z2b}.
Notably, among all the models, the \textbf{z0b} model exhibits the highest average net accretion rate, primarily driven by its steady gas inflow and relatively weak outflows.

% \begin{figure}
%     \centering
%     \begin{tabular}{c}
%     %\centering
%         \includegraphics[width=0.5\textwidth]{ACCR_baryon_0.2vir.png} 
%         \includegraphics[width=0.5\textwidth]{ACCR_baryon_1vir.png} 
%     \end{tabular}
%     \caption{ The evolution of mass accretion rates during the simulation. We calculate the mass accretion rate by temporal variation of the enclosed mass of galactic disk scale within $< 0.2 \, R_\mathrm{vir}$ (left) and of a halo scale of $R_\mathrm{vir}$ (right). This mass accretion includes both mass inflow and outflow of the targeted scale. Therefore, the positive accretion rate means the inflow mass is larger than the outflow, and the negative accretion rate shows the inflow mass is smaller than the outflow. For all models, the accretion rate at a galactic disk scale of $0.2 \, R_\mathrm{vir}$ is overall positive, indicating the growth of the galaxy with continuous gas fueling. In contrast, the mass accretion rate at $R_\mathrm{vir}$ can be negative due to strong outflows. The accretion of $z = 1$ and $z = 2$ models demonstrate significant variation in both galactic disk-scale and halo-scale. }
%     \label{fig:accretion_old}
% \end{figure}
%  
%

\subsection{Metallicity of Galaxies and Their CGM}
\label{subsec:metallicity}

\begin{figure}
    \centering
    \begin{tabular}{ccc}
            \includegraphics[width=1\textwidth]{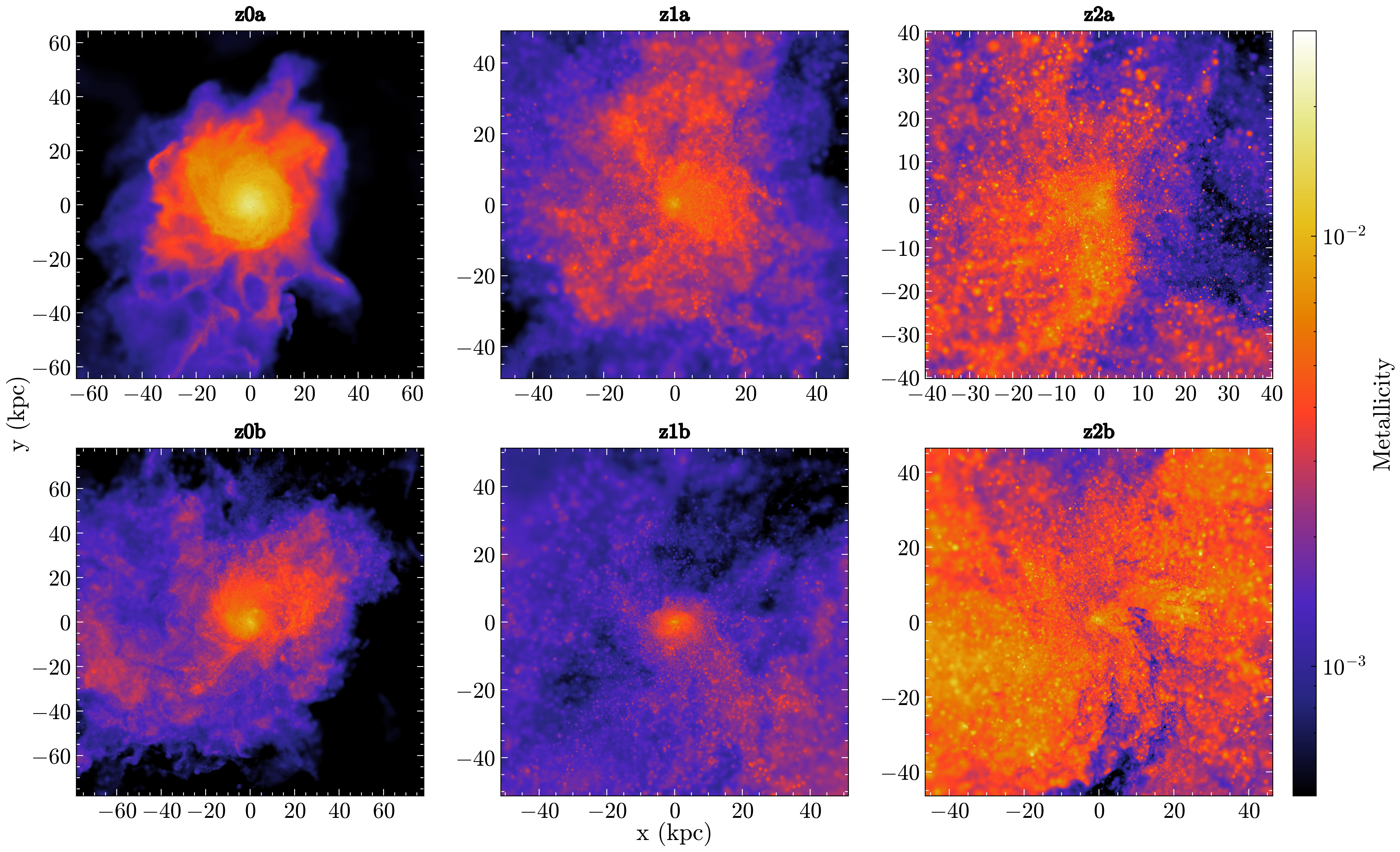}
            \end{tabular}
    \caption{The metallicity distribution of halo gas at the end of the simulations.  The metal distribution in $z = 2$ halos is more extensive than that of $z = 0$ and $z = 1$ halos.  }
    \label{fig:metal}
\end{figure}

We examine the chemical properties of DGs and their CGMs by presenting their absolute metallicity \footnote{The absolute metallicity of our Sun ($Z_\odot$) is $\sim 0.02$.} distribution in Figure \ref{fig:metal}.
In \textbf{z0a}, the metal-rich gas of the galaxy is smoothly mixed and diffused into its CGM.
The CGM at $z = 0$ shows a distinct drop in metallicity in the outer regions, consistent with observations \citep{Bordoloi2014}.  
However, for the other redshifts, the metal is widely distributed throughout the entire CGM, implying strong galactic outflows that chemically enrich the CGM.  
Furthermore, the CGM in the $z = 2$ models shows clumpy structures in the metal distribution due to the mixing of metal-rich gas ejected from the central galaxies and metal-poor gas accreted from IGM.
Some ejected gas clumps, containing metal-rich supernovae ejecta, can reach a higher metallicity than the disk gas \citep{2018Christensen}.

\begin{figure}
    \centering
    \begin{tabular}{ccc}
        \includegraphics[width=1\textwidth]{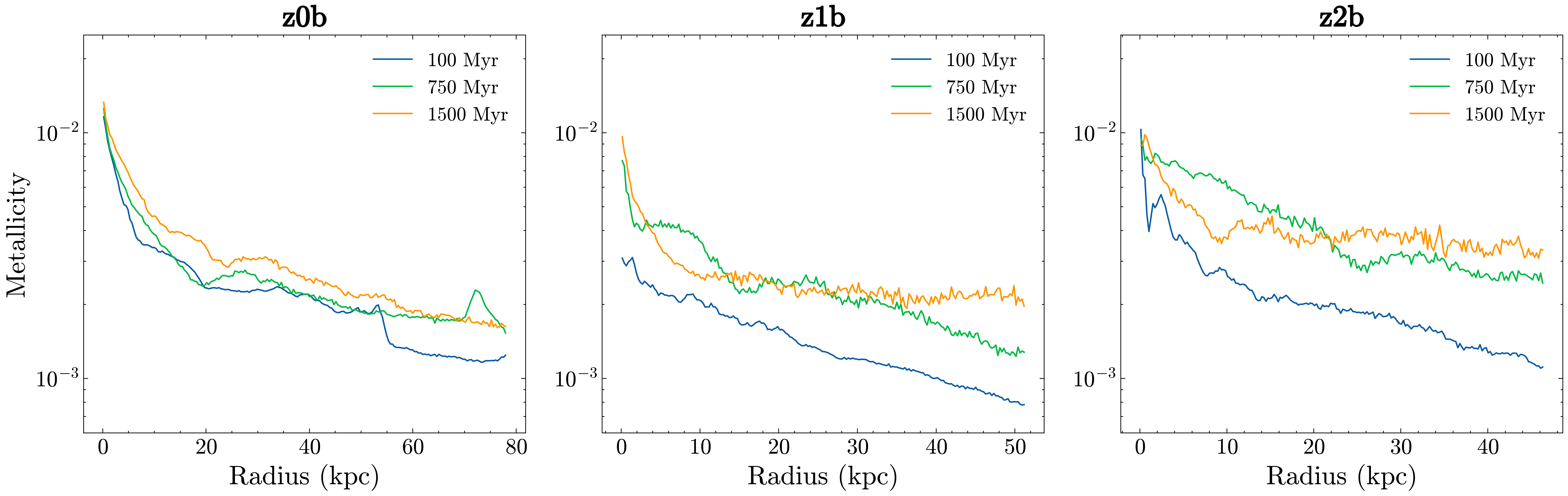} 
    \end{tabular}
    \caption{Radial metallicity of halos. The peak metallicity appears in the halo center below the solar metallicity of $0.02$. The increasing CGM metallicity is from the metal produced in the star-forming region that is further transported to the entire halo by supernovae and AGNs.}
    \label{fig:metal_dist}
\end{figure}

We show the evolution of the 1D metallicity profile in Figure \ref{fig:metal_dist}.
In the evolution of the metallicity profiles, the overall metallicity of \textbf{z0b} gradually increases over time.  
The metallicity profiles of $z = 1$ and $z = 2$ eventually become nearly uniform at $r > 10$ pc, suggesting efficient metal mixing in the CGM.  
The shallower gravitational potential of DGs allows them to eject more metal-rich gas from supernovae into the CGM than more massive galaxies \citep{2018Christensen}.
At the end of the simulations, the metallicity at the halo boundary increases by a factor of $2-3$ from the beginning. 
However, in \textbf{z1b} and \textbf{z2b}, the metallicity distribution decreases within 10-20 kpc from 750 Myr to 1500 Myr due to the strong outflow driven by stellar and AGN feedback that transport the inner metal outward (see Figure \ref{fig:accretion}). 

The average metallicity of most DGs increases 30\% and their corresponding total metal mass increases by a factor of 1.5–3 at the end of the simulation. Except, the metallicity of \textbf{z0a} decreases by about 14\% due to the dilution by accretion of metal-poor gas and weaker SF activity.
On the other hand, \textbf{z2b} shows the smallest increase of total metal mass by $\sim 10\%$ due to a strong stellar feedback that ejects much newly synthesized metal.

% , even though the continuous inflow of pristine gas that dilutes the metallicity
% and the strong inflow of pristine gas from surrounding filaments

% dark matter
\subsection{Dark Matter Mass and Structure}
\label{subsec:dark matter}

\begin{figure}
    \centering
    \includegraphics[width=1\textwidth]{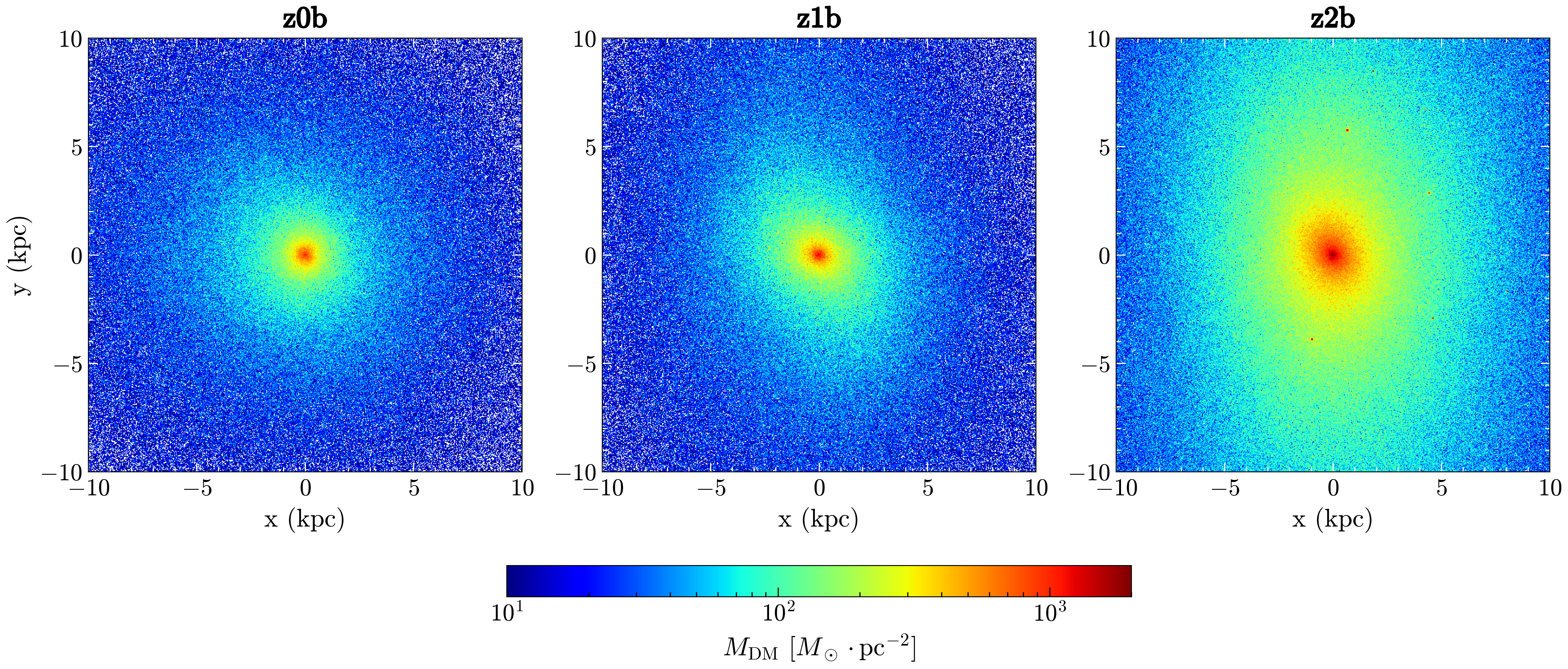}
    \caption{The inner dark matter structure of halos at the end of the simulations. These structures look spherical, and their radial profiles follow the NFW profiles \citep{1996Navarro}. In \textbf{z2b}, several little red dots are scattered around the outskirts of the halo center possibly due to the minor merger of the surrounding mini halos. }
    \label{fig:dm_dist}
\end{figure}

To study the dynamics of galaxies, we compare the distribution of dark matter for all models in Figure \ref{fig:dm_dist}. The structure of the dark-matter halo varies between redshifts. The halo exhibits a virialized spherical distribution in \textbf{z0b}.
In contrast, the dark matter halos of \textbf{z1b} and \textbf{z2b} are more elongated.
Moreover, several dense clusters of dark matter (scattered little red dots in \textbf{z2b} of Figure \ref{fig:dm_dist}) orbit around the halo center in \textbf{z2b}, suggesting minor mergers of tiny dark matter halos. 
These mergers may influence the evolution of the host galaxy via their tidal interactions.
\\

% 5. The AGN feedback from CGM accretion: AGN accretion rate
\subsection{The Growth of Supermassive Black Holes}
\label{subsec:smbh}
\begin{figure}
    \centering
    \includegraphics[width=0.65\textwidth]{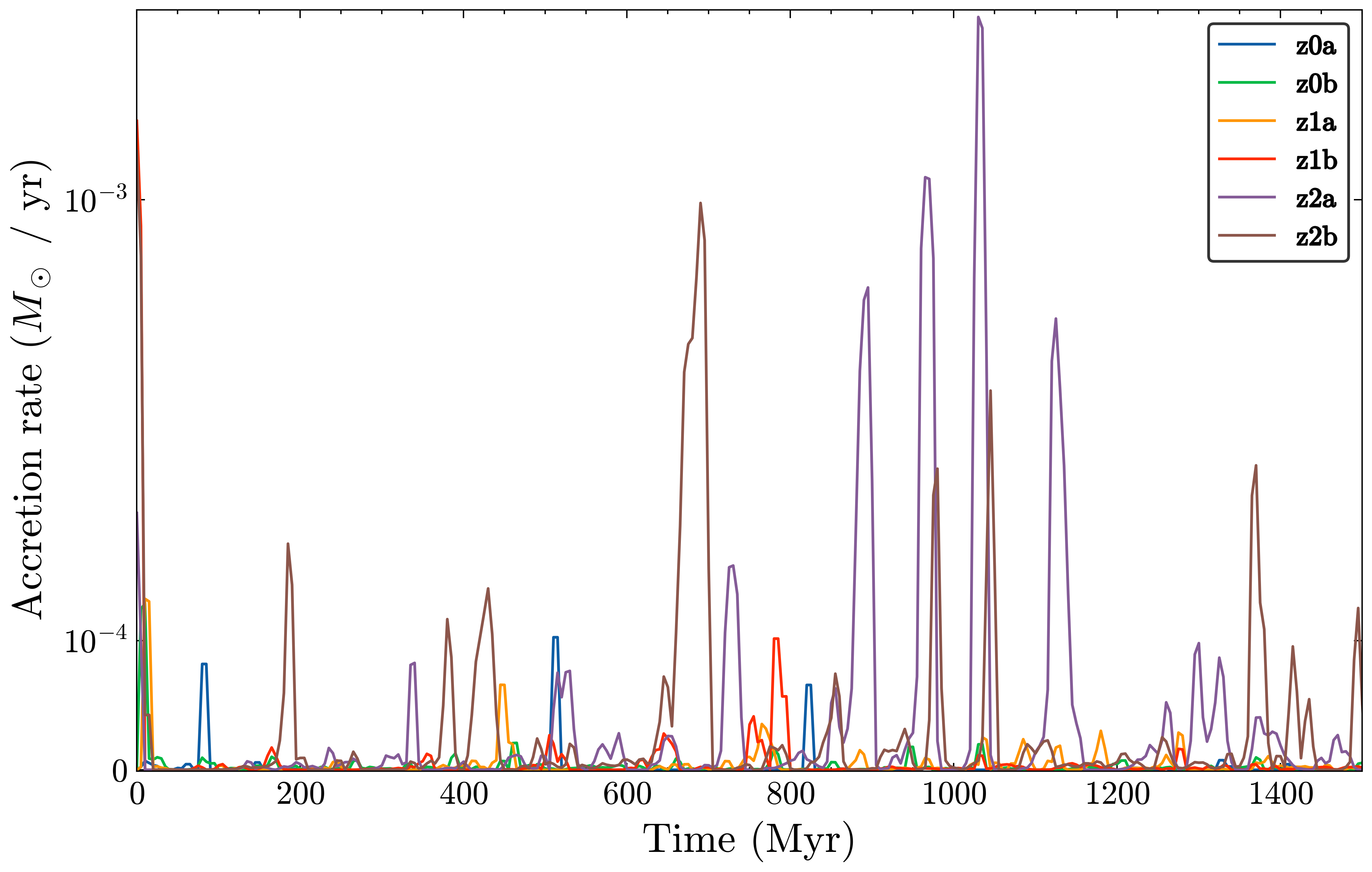}
    \caption{The accretion history of SMBH throughout the simulation. The accretion histories show a bursty pattern for $z=1$ and $z=2$ models, suggesting several duty cycles of AGN activity and the rapid growth of SMBHs. The peak accretion rates reach $\sim 10\%$ Eddington rates.   }
    \label{fig:smbh_acc}
\end{figure}

To evaluate the impact of SMBH on DGs, we show the history of accretion rate on SMBH in Figure \ref{fig:smbh_acc}. Most of the SMBHs exhibit quasiperiodic outbursts in the accretion rate history, implying a short duration of the active duty cycle. The highest SMBH accretion rate occurs in \textbf{z2a} at 1000 Myr surrounded by several smaller bursts with a time separation of $\sim 100$ Myr. The peak rates of \textbf{z2a} and \textbf{z2b} can reach $\sim 10\%$ of the Eddington accretion rate,
$$\dot{M}_\mathrm{Edd} = 0.022 \, M_{\text{6,BH}} \quad \Ms\ \text{yr}^{-1} $$
, where $M_{\text{6,BH}}$ is the mass of SMBH in units of $10^6 \, M_\odot$.
This AGN activity also leads to intense feedback to CGM by blowing the metal-rich gas away from the galaxy into CGM and creating a strong outflow. 

As discussed in section \ref{subsec:gas accretion}, the galaxies' high gas accretion rate possibly triggers the busty accretion of the SMBH.  
For \textbf{z2b}, the SMBH accretion is primarily driven by the gas accretion from filamentary CGM structures. 
Meanwhile, part of the accretion in \textbf{z2a} is from the minor mergers of smaller surrounding structures.  
For the gas-rich galaxy such as \textbf{z0a}, the SMBH activity can be triggered even without a high accretion rate from CGM. 
Compared with \textbf{z0a} and \textbf{z0b}, the SMBH accretion rate in \textbf{z0b} is lower despite its higher CGM accretion rate.

% star formation
\subsection{Star Formation}
\label{subsec:SF}

\begin{figure}
    \centering
    \begin{tabular}{c}
    %\centering
        \includegraphics[width=0.65\textwidth]{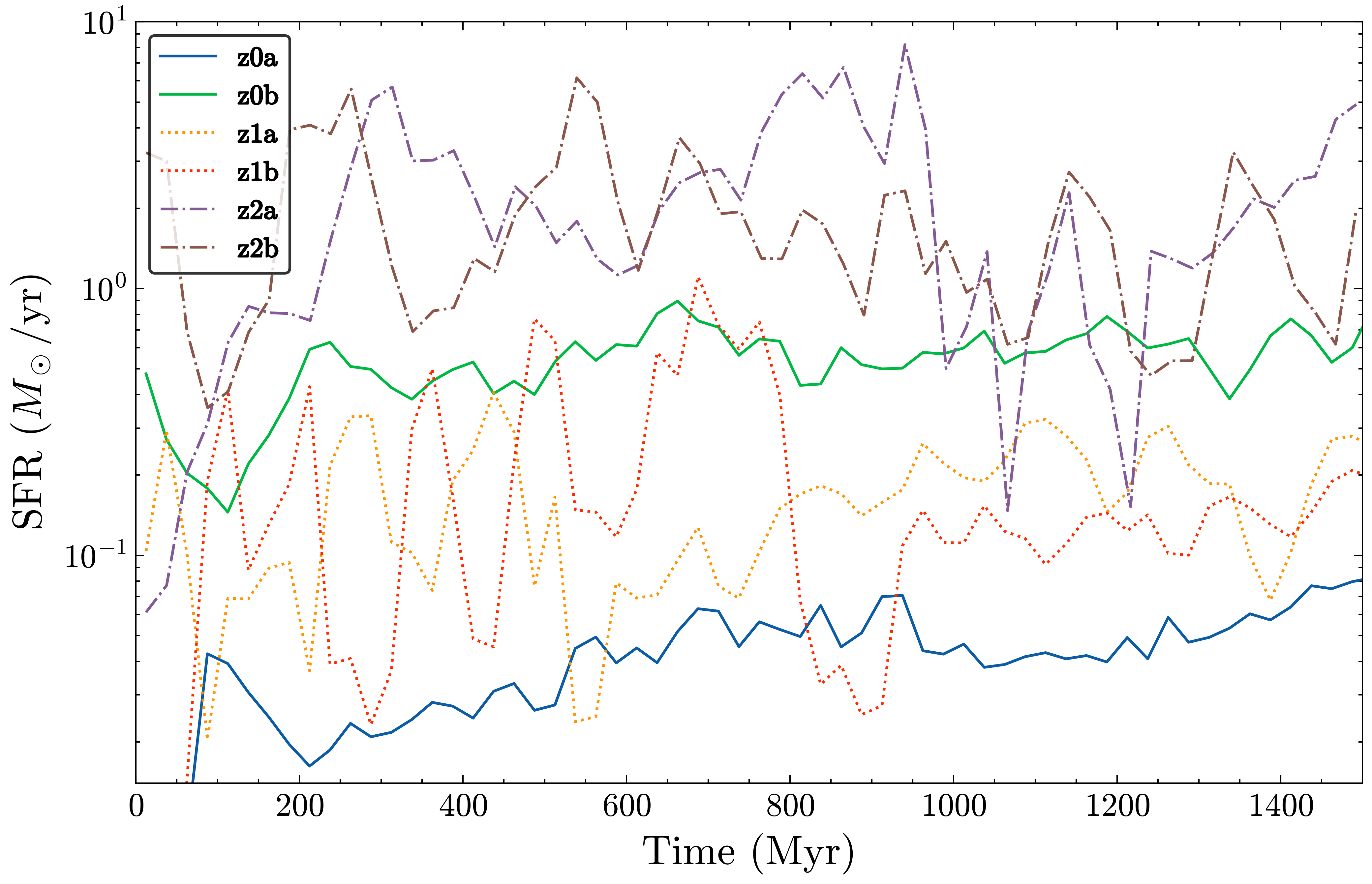} 
    \end{tabular}
    \caption{ Star formation rate histories for all models. DGs at $z=1$ and $z=2$ show bursty SFH with variations spanning a range of 2 dex, which is also found in \cite{2014Shen}. As $z$ increases, the average SFR in DGs also increases.}
    \label{fig:sfr}
\end{figure}

% star formation rate (SFR)
We show the evolution of SFRs of all models in Figure \ref{fig:sfr}. In general, the gas accretion rate from CGM is sufficient to sustain the SFR for galaxies among all redshifts. However, minor mergers, stellar, and SMBH feedback also affect the galactic SFR and are reflected in the gas accretion rates, as shown in Figure \ref{fig:accretion}. For \textbf{z0a} and \textbf{z0b}, the evolutionary track of SFR is steady and increases slightly during simulation.  
Moreover, the more compact stellar disk can survive under the strong SMBH feedback and accretion flow, as shown in Figures   \ref{fig:accretion} and \ref{fig:smbh_acc}.

For \textbf{z1a} and \textbf{z1b}, their average SFR is lower than that of \textbf{z0b}.  
Because the gas accretion of \textbf{z1a} and \textbf{z1b} are subject to tidal interactions from their distant neighbors of massive galaxies that affect the CGM and surrounding IGM, it causes a drop in the accretion rate of CGM and IGM.  
The SFR also exhibits larger variability during the first 100 Myr, then becomes more steady, which correlates well with the gas accretion rate around the galactic scale.  
This behavior is also related to the rapid-evolving structure of the gas disk and dark matter halo. 

For galaxies at $z = 2$, their SFRs are $10-100$ higher than those of galaxies at $z=0$ and $z=1$.  
The high accretion rate and irregular shape of the galaxy allow cold gas to flow directly into the center, boasting SFRs while simultaneously increasing outflows due to feedback from stars and the SMBH, leading to a rapid variation in SFR.  
The dynamic impact of strong inflows from CGM and minor mergers can also alter the structure of the galactic disk and its SFR.  

\begin{figure}
    \centering
    \begin{tabular}{c}
    %\centering
        \includegraphics[width=1\textwidth]{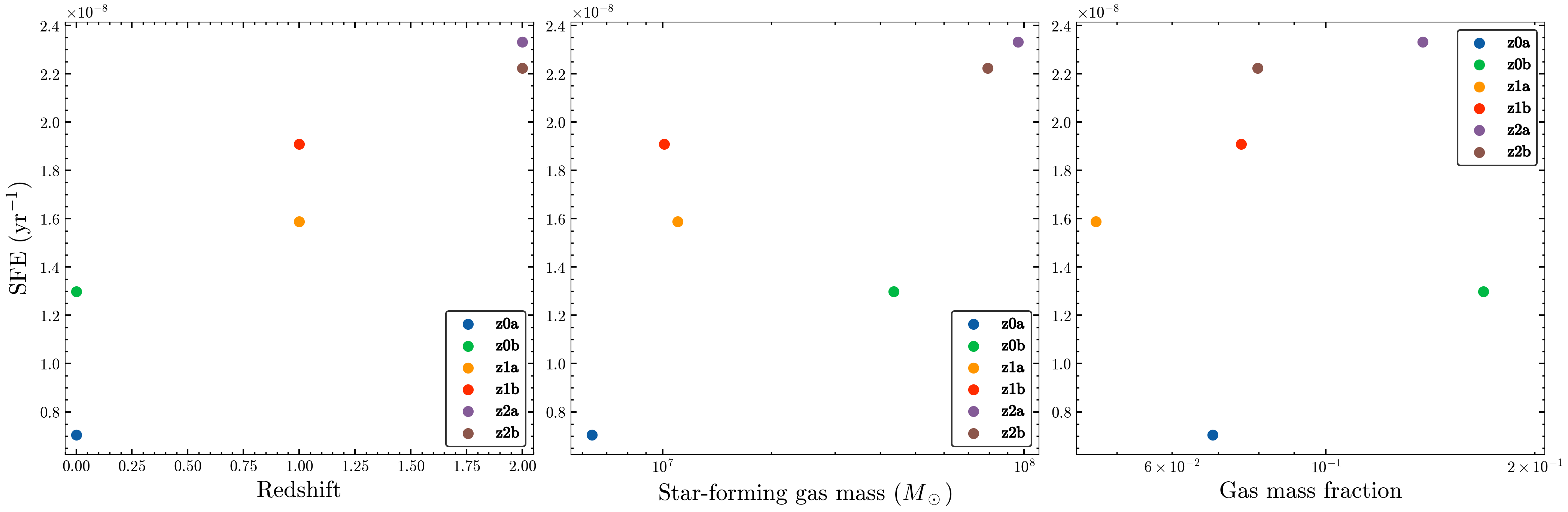} 
    \end{tabular}
    \caption{The SFEs with respect to redshift (left), star-forming gas mass (middle), and gas mass fraction within DG halos (right). The amount of star-forming gas mass is to sum the molecular mass of gas density of $>0.1$ 
 cm$^{-3}$. SFEs have a stronger correlation with redshift than the amount of star-forming gas mass and gas mass fraction inside the halo.}
    \label{fig:sfe}
\end{figure}

% SFE
We further examine the correlation between the star-formation efficiency (SFE) and other physical quantities in Figure \ref{fig:sfe}. SFE is mostly sensitive to the redshift than the amount of star-forming mass and molecular gas fraction of DGs. The left panel in Figure \ref{fig:sfe} shows that SFE increases as $z$ increases and the \texttt{SIMBA} simulations \citep{2019Dave, 2024Ghodsi} also found similar results. The middle and right panels of Figure \ref{fig:sfe} show weak correlations between SFE and star-forming gas mass and fraction. This result demonstrates $z$ has a stronger impact than the halo properties of DGs on their SFE.  

%, which also reflects the redshift-driven changes in gas density distributions shown in Figure \ref{fig:phase}. In contrast, the halo environment exhibits little or no clear correlation with the SFE, suggesting that the SFE is only weakly dependent on halo properties.
%}

% The dispersion is higher at lower redshift, even though the DGs appear more scattered at $z = 2$ in Figure \ref{fig:star_gas_tng}.% Although the environment influences galaxy evolution, it plays only a minor role in mass contribution, as discussed in Section 4.2. 

% newborn
\subsubsection{Spatial Distribution of SF Region}
\begin{figure}
    \centering
    \includegraphics[width=1\textwidth]{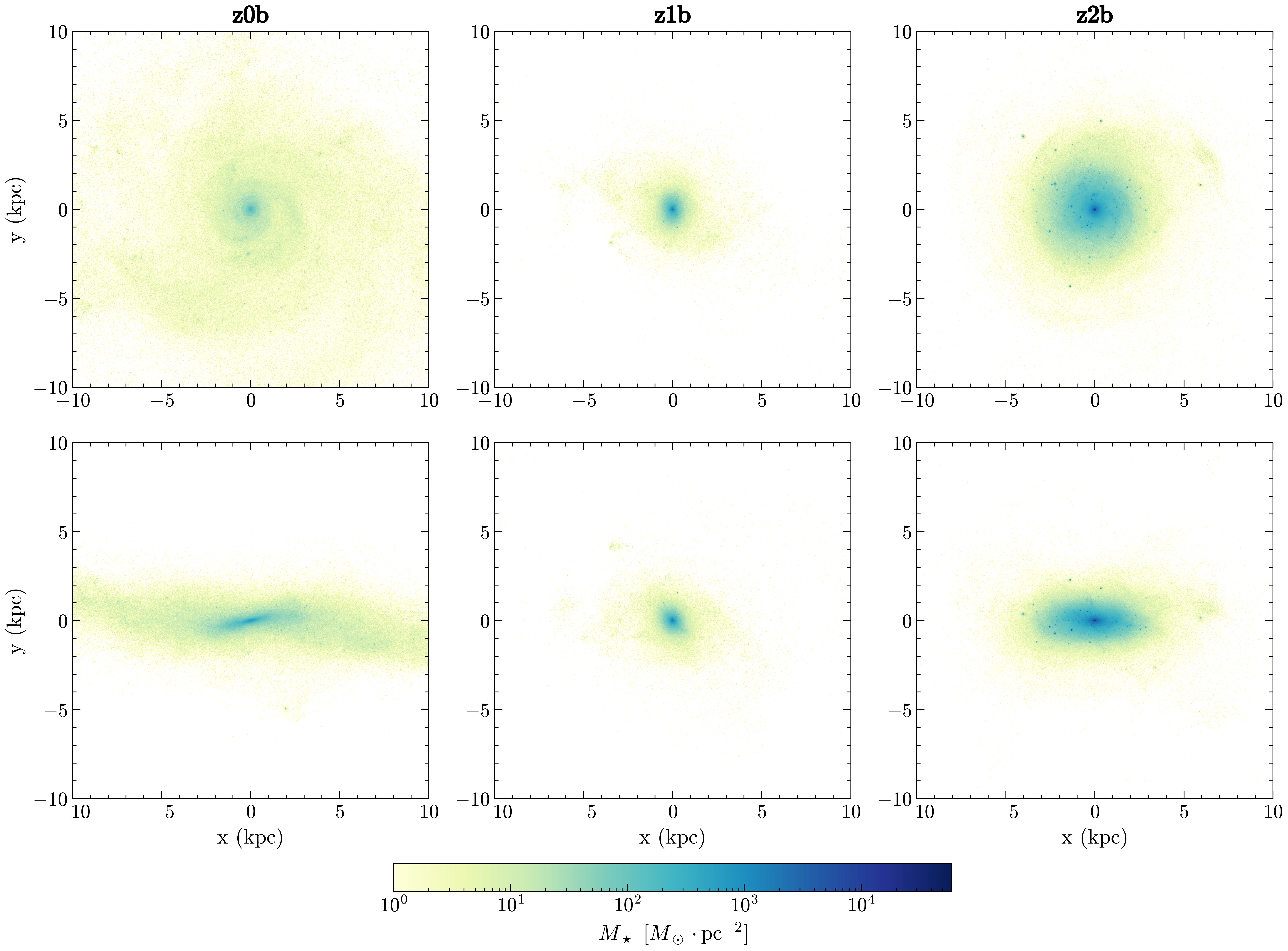}
    \caption{Spatial distribution of stars formed within the inner region of 5 kpc at the end of simulation. The top and bottom panels show the face-on and edge-on view of galaxies, respectively. Among these models, only \textbf{z0b} shows a spiral structure. In \textbf{z2b}, some scattering star formation regions appear in the outskirt of the disk due to the star formation in the clumpy disk. This clumpy structure may be driven by disk instabilities via the strong gas accretion from CGM.}

    \label{fig:star_dist}
\end{figure}

We show the stellar-disk structure of DGs in Figure \ref{fig:star_dist}. The SF regions become more compact both in the gas disk and in the bulge as $z$ increases.  
In \textbf{z1b}, due to the unstable structure and pulsational outflows of the galactic disk, the SF region appears to be more elliptical or bulge-dominated.  
In \textbf{z2b}, with continuous gas supply and high gas density, the SF region is populated throughout the disk with the peak at the center. The clumpiness of the SF density is likely generated by the disk instability via strong accretion.  
Combined with the SMBH accretion rate in section \ref{subsec:smbh}, both explain why the strong accretion flow on the galactic scale at $z = 2$ can effectively reach the galactic center, triggering both active SF and SMBH activity.

\section{Discussion}

\subsection{The Metal Enrichment of the CGM}
\label{subsec:metal_enrich}
\begin{figure}
    \centering
    \includegraphics[width=1\textwidth]{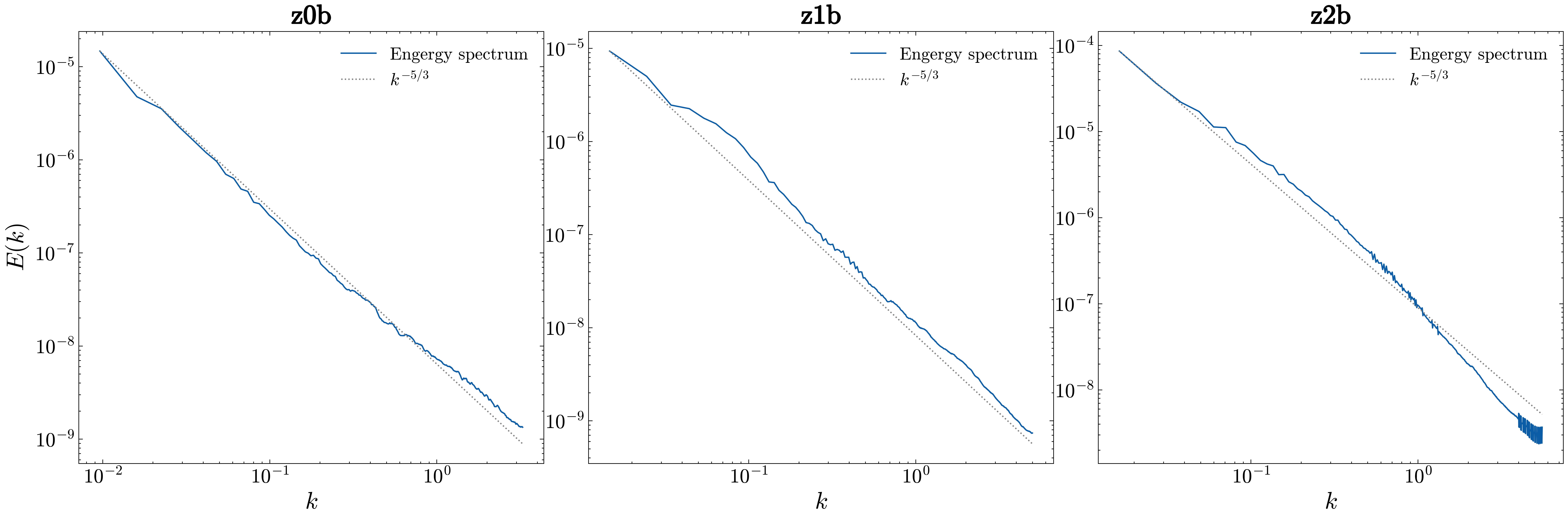} 
    \caption{Kinetic energy power spectrum of the halo gas at the end of the simulation. The slope of gas spectra matches well with the dotted line, presenting the slope of $-5/3$ for the typical Kolmogorov spectrum. These profiles suggest the entire halo gas is highly turbulent. }
    \label{fig:kolmogorov}
\end{figure}

Cooling of hot gas by metals through line emission is efficient. Therefore, the metal content of the CGM can affect the evolution of the galaxy and its CGM. 
Furthermore, the metals ejected from the galaxy can serve as a tracer in the CGM to probe the coevolution of the galaxy and the CGM.
In Section \ref{subsec:metallicity}, we examine the spatial distribution of metallicity and its temporal evolution.
To evaluate the mixing within the simulation box, we present the kinetic energy spectra of the gas in Figure \ref{fig:kolmogorov} based on the snapshots in Figure \ref{fig:metal_dist}.
For all halos, the slope of the power spectra is $-5/3$, aligning well with the Kolmogorov spectra \citep{Zakharov1992, Scalo2004}.
This result suggests that the gas within the halo scale is highly turbulent.

Large-scale mixing is driven by interactions between halo inflows and galactic outflows. This mixing can disperse metal from galaxies to CGM efficiently. Furthermore, the gas inflow rates determine the star formation and consequent stellar feedback. Therefore, the inflow influences the total metal mass within the galaxies.
For z2b, the slope of the spectrum becomes slightly steeper than $-5/3$ at large $k$, indicating weaker turbulence on the small scale. This may result in insufficient mixing of gas clumps ejected from the DGs, as shown in Figure \ref{fig:metal}, in contrast to the well-mixed CGM at lower redshifts galaxies suggested by \cite{2019Hafen}.
% For \textbf{z2b}, the slope of the spectrum becomes slightly steeper than $-5/3$ in large $k$, indicating weaker turbulence on the small scale that may cause the clumpy structure found in Figure \ref{fig:metal}, \mynote{as the well-mixed CGM compared to lower redshifts galaxies suggested by \cite{2019Hafen}.}
However, strong stellar and SMBH feedback can still transport metals in \textbf{z2a} and \textbf{z2b} to the halo boundary or beyond, as shown in the metallicity distribution profiles of Figure \ref{fig:metal_dist}.
% This result agrees with \cite{2020Hafen} that most of the CGM gas is ejected into the IGM for DG halos at z = 2.
%\mynote{This result agrees with \cite{2020Hafen} that most of the CGM gas of z = 2 DG halos is ejected into the IGM by z = 0 but replenished by
%}

% 2. The redistribution of CGM and IGM gas?
\subsection{The Impact of Evolved CGM and IGM Gas on Host Galaxies}
We now quantify the mass of accreting gas during the evolution and discuss its impact on the SFR and the gas mass evolution of galaxy and halo. 
By tracing the gas in the simulations, we can follow the gas evolution of galaxy, CGM, and IGM, as well as their contributions to star formation.
We label each gas particle at $100$ Myr to allow some relaxation time after splitting the gas particles. We then track these particles until the end of the simulation at $1500$ Myr. Furthermore, we can calculate the mass change rates for gas originated from the galaxy, CGM, and IGM separately, unlike the total mass change rate of the enclosed mass with a given region, as shown in Figure \ref{fig:accretion}.
\\
\\

\begin{figure}
    \centering
    \includegraphics[width=1\textwidth]{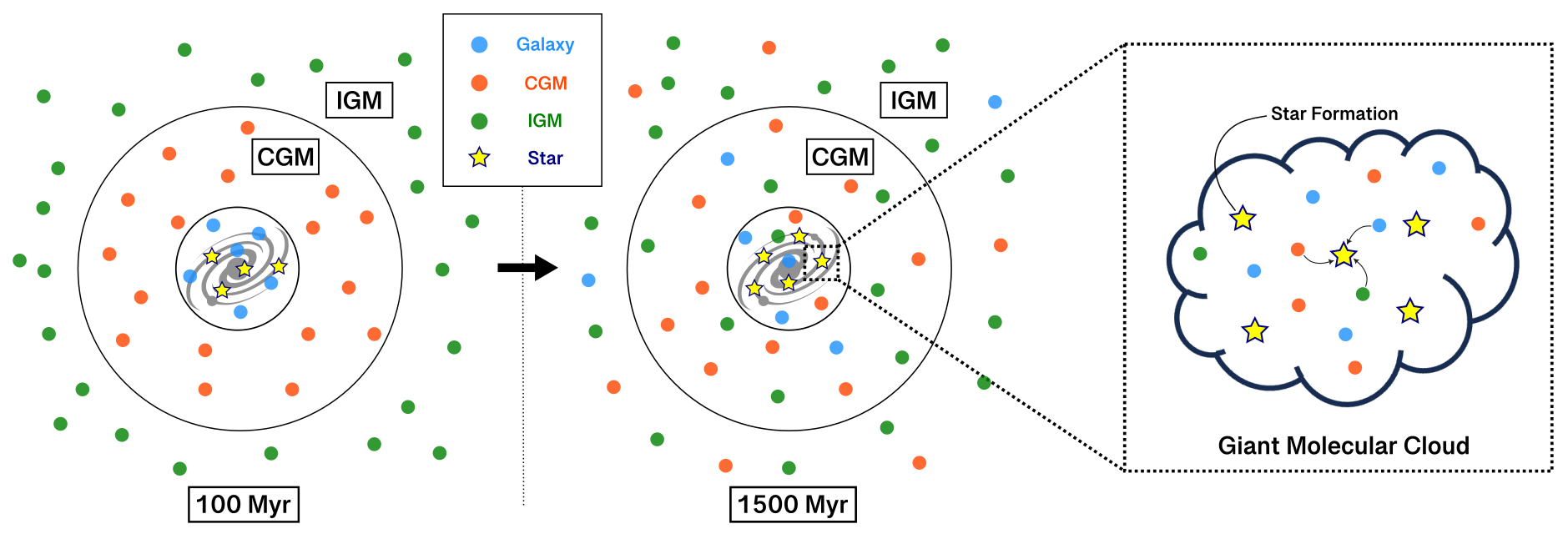}
    \caption{Schematic of gas particle tracing. Blue, orange, and green dots represent the gas particles initially from galaxy, CGM, and IGM at the beginning of the simulations, respectively. The middle panel shows the gas distribution at the end of simulation and color dots are mixed due to the gas accretion and outflow. The right panel shows the close-up of the star formation region within the galaxy. Gas accreted from CGM and IGM partially contribute to the star formation within the galaxy.  }
    \label{fig:schematic_trace}
\end{figure}

\begin{figure}
    \centering
    \begin{tabular}{ccc}
    %\centering
        \includegraphics[width=0.33\textwidth]{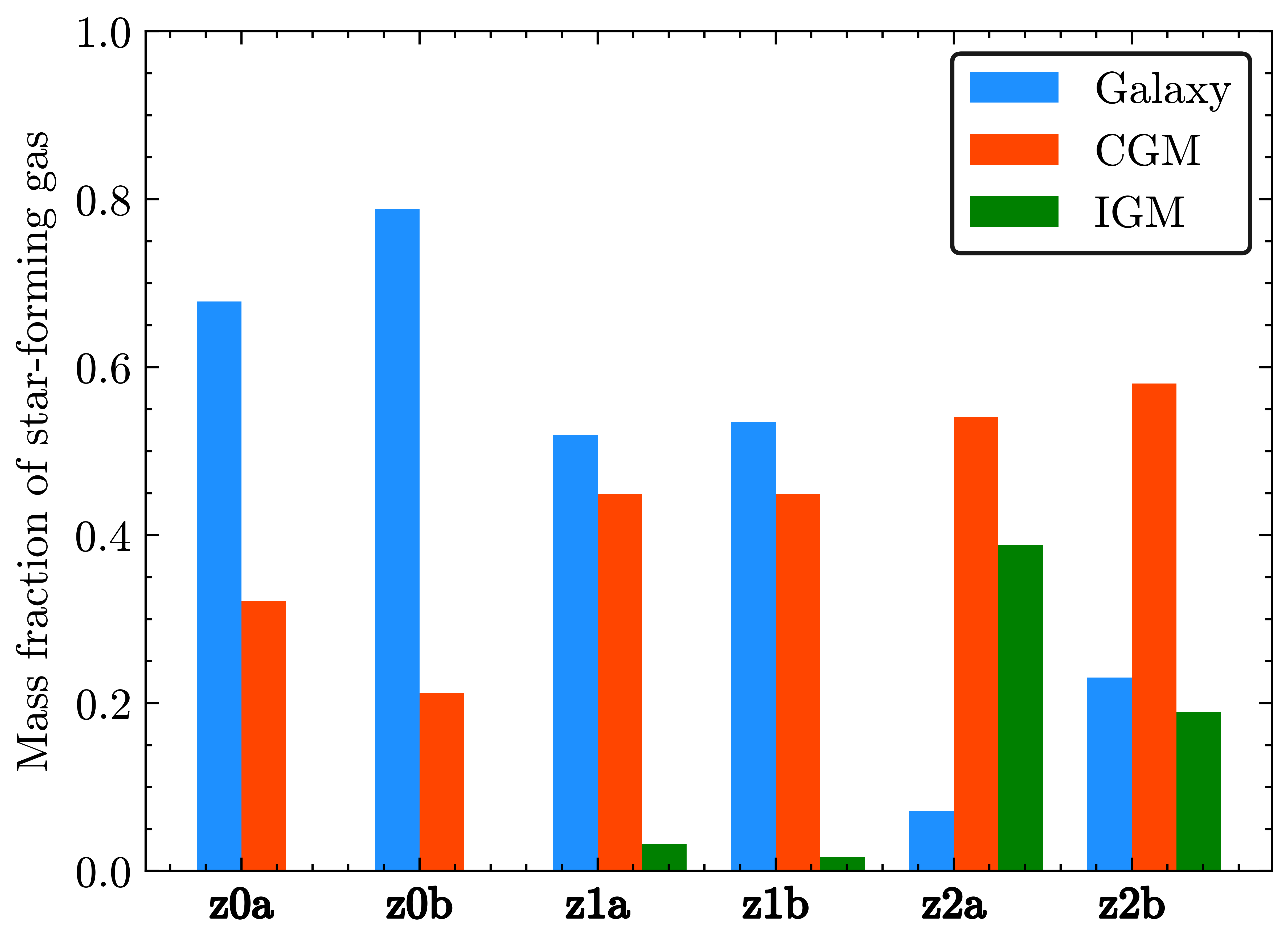} & 
        \includegraphics[width=0.33\textwidth]{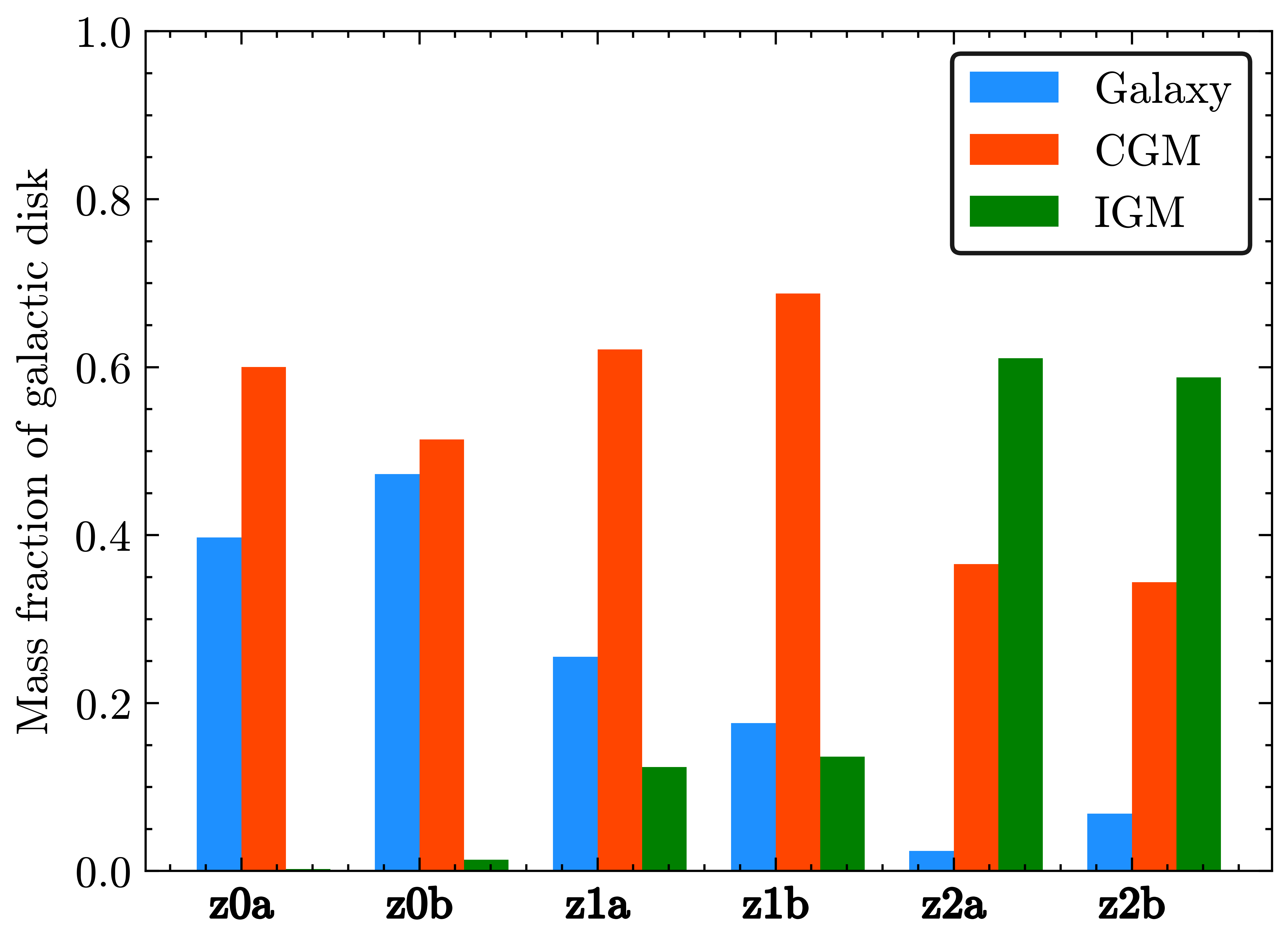} &
        \includegraphics[width=0.33\textwidth]{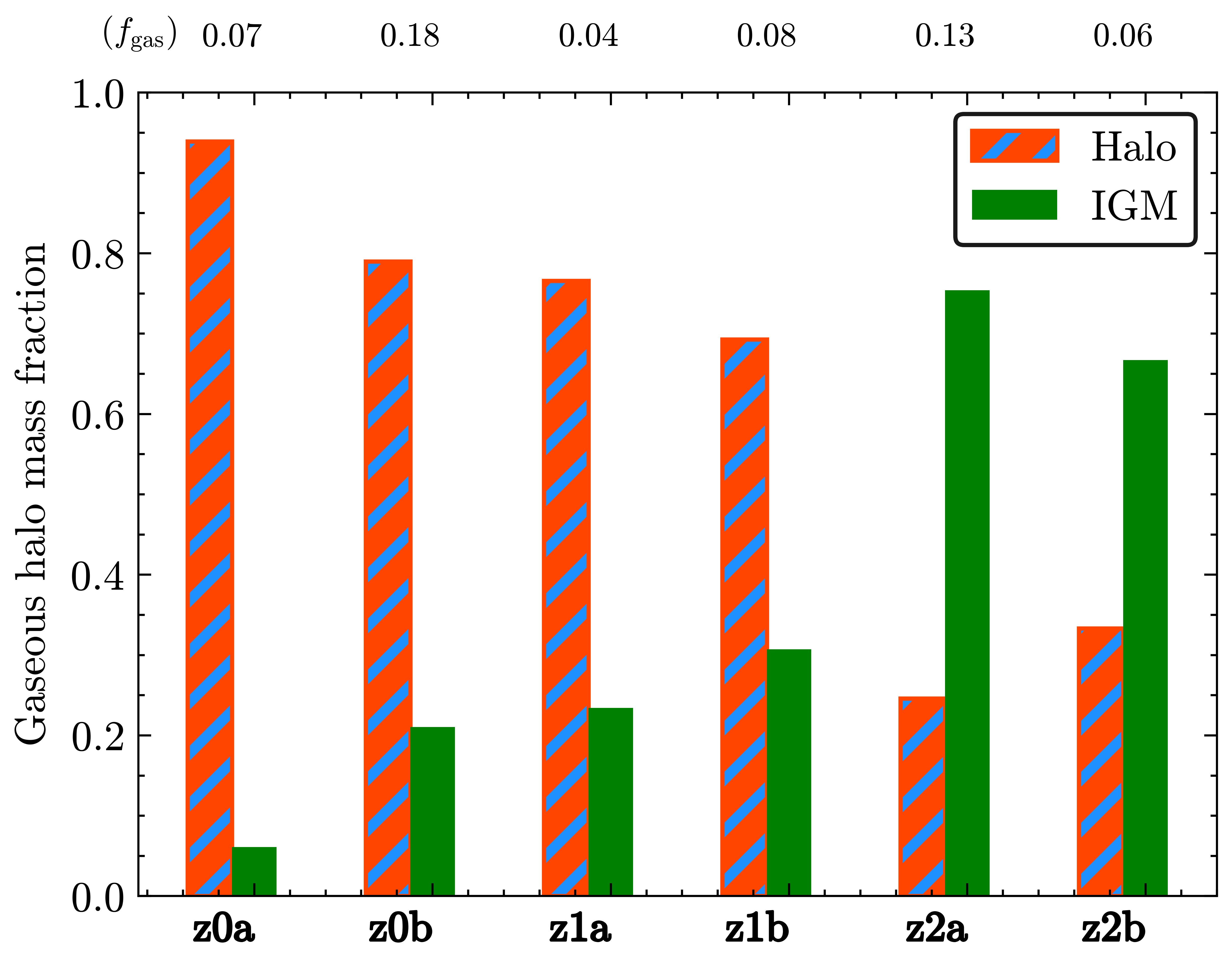} \\         
    \end{tabular}
    \caption{Gas contribution from the original galaxy, CGM, and IGM to the star-forming gas (left), galactic disk (middle), and halo gas (right) at the end of the simulation. The orange bars with blue stripes in the right panel represent the total gas mass inside a halo, including galaxy and CGM gas. These contributions are calculated using a particle tracking technique by tagging each gas particle and following its trajectory till its final destination. Gas mass fraction of the entire galactic halo ($f_\mathrm{gas}$) is added on top of the gaseous halo plot.}
    \label{fig:mass_contribution}
\end{figure}
% Although $f_\mathrm{gas}$ does not directly correlate with the mass fraction from the IGM,
%  At a given $z$, halos with a higher IGM   show higher $f_\mathrm{gas}$.

\begin{deluxetable*}{c||cccc|ccc|c}
\tablecaption{The averaged rates of star-forming gas ($\Dot{M}_{\star}$), disk gas ($\Dot{M}_\text{Disk}$), and halo gas ($\Dot{M}_\text{Halo}$) contributed by the initial galaxy, CGM, and IGM.  \label{tb:acc_SFR}}
\tablewidth{0pt}
\tablehead{
Model & \multicolumn{4}{c|}{$\Dot{M}_{\star}$ [$M_\odot \cdot \text{yr}^{-1}$]} & \multicolumn{3}{c|}{$\Dot{M}_\text{Disk}$ [$M_\odot \cdot \text{yr}^{-1}$]} & $\Dot{M}_\text{Halo}$ [$M_\odot \cdot \text{yr}^{-1}$] \\
name & (Galaxy) & (CGM) & (IGM) & (Total) & (CGM) & (IGM) & (Total) & (IGM)
}
\startdata
\textbf{z0a} & $0.026$ & $0.012$ & $0.0$ & $0.038$ & $0.630$ & $0.002$ & $0.632$ & $0.081$ \\
\textbf{z0b} & $0.380$ & $0.102$ & $0.0$ & $0.482$ & $1.843$ & $0.048$ & $1.891$ & $1.360$ \\
\tableline
\textbf{z1a} & $0.076$ & $0.066$ & $0.005$ & $0.147$ & $0.373$ & $0.074$ & $0.447$ & $0.252$ \\
\textbf{z1b} & $0.106$ & $0.089$ & $0.003$ & $0.198$ & $0.724$ & $0.143$ & $0.867$ & $0.630$ \\
\tableline
\textbf{z2a} & $0.158$ & $1.194$ & $0.856$ & $2.208$ & $0.887$ & $1.482$ & $2.369$ & $5.562$ \\
\textbf{z2b} & $0.385$ & $0.971$ & $0.316$ & $1.672$ & $0.409$ & $0.700$ & $1.109$ & $2.920$ \\
\enddata
\end{deluxetable*}

% for star formation
\subsubsection{The Star Formation}

The left panel of Figure \ref{fig:mass_contribution} shows the contribution of a star-forming gas from the galaxy, CGM, and IGM.
In general, the contributions of the galaxy, CGM, and IGM depend on $z$. The original gas residing in the galaxy accounts for $\sim 70\%$ star-forming gas for galaxies at $z = 0$ and $\sim 50\%$ for galaxies at $z= 1$. The value drops further as the redshift increases. 
At $z = 2$, the gas originating from the galaxy contributes only $< 20\%$ to the star formation gas. 
Strong accretion and outflow for high-$z$ galaxies cause the initial galactic gas to become a subdominant component in the SF. 

Meanwhile, the gas contribution to the SFR from CGM and IGM increases with redshift. The accreted CGM gas dominates the mass in the reservoir of star formation from 20\% of gas for $z = 0$ galaxies to 50\% for $z = 2$ galaxies.
For the contribution from IGM, it is much more sensitive to the redshift, from 0. 01\% to 20\%, when $z$ increases from 0 to 2.
The influence of IGM is minor for galaxies at $z = 0 $ and $ 1$. 
However, the evolution of $z>2$ galaxies should include their IGM. 
In sum, our results demonstrate the importance of CGM and IGM for the evolution of high-$z$ galaxies.

% for total mass in the disk
\subsubsection{Mass and Structure of Galaxy}

In the middle panel of Figure \ref{fig:mass_contribution}, we present the contribution of gas inside the galaxy, defined as the gas with a distance from the central SMBH of less than $0.2 \ R_{\text{vir}}$.
The result is different from the contribution of star formation.
During the $1.5$ Gyr simulation, the CGM and IGM gas can eventually contribute $> 50\%$ of the gas inside the galaxy, even for \textbf{z0a}, having minimal environmental accretion beyond the halo scale.
This suggests that CGM plays a critical role in the galaxy's evolution across cosmic time.
The contribution of gas from the CGM can account for around $40-70\%$ of the total gas mass in the galaxy. Furthermore, it should be mentioned that the effect of CGM is more important at $z = 1$ (70\%). For the contribution of IGM to the galactic gas, the percentage increases from $z = 0$ to $z = 2$ and reaches $60\%$ at $z = 2$, surpassing CGM.

% for total mass in the halo
\subsubsection{Halo Mass and Size}

In the right panel of Figure \ref{fig:mass_contribution}, we present the contribution of the total gas mass inside the halo of $<R_{\text{vir}}$. 
The mass fraction contributed by IGM to the halo mass differs between redshifts.
For $z = 0$ and $z = 1$, except for the isolated galaxy \textbf{z0a}, IGM can contribute $20\%-30\%$ of the total gas mass inside the halo after 1.5 Gyr of evolution.
For the $z = 2$ galaxies, the contribution of IGM can account for $60\%-70\%$ of the gas in the halo, dominating the entire gas reservoir over the galaxies and CGM in the early universe.

% comparison
\subsection{Comparison to Previous Studies and Observations}
\label{subsec:comparison}

\begin{figure}
    \centering
    \includegraphics[width=0.65\textwidth]{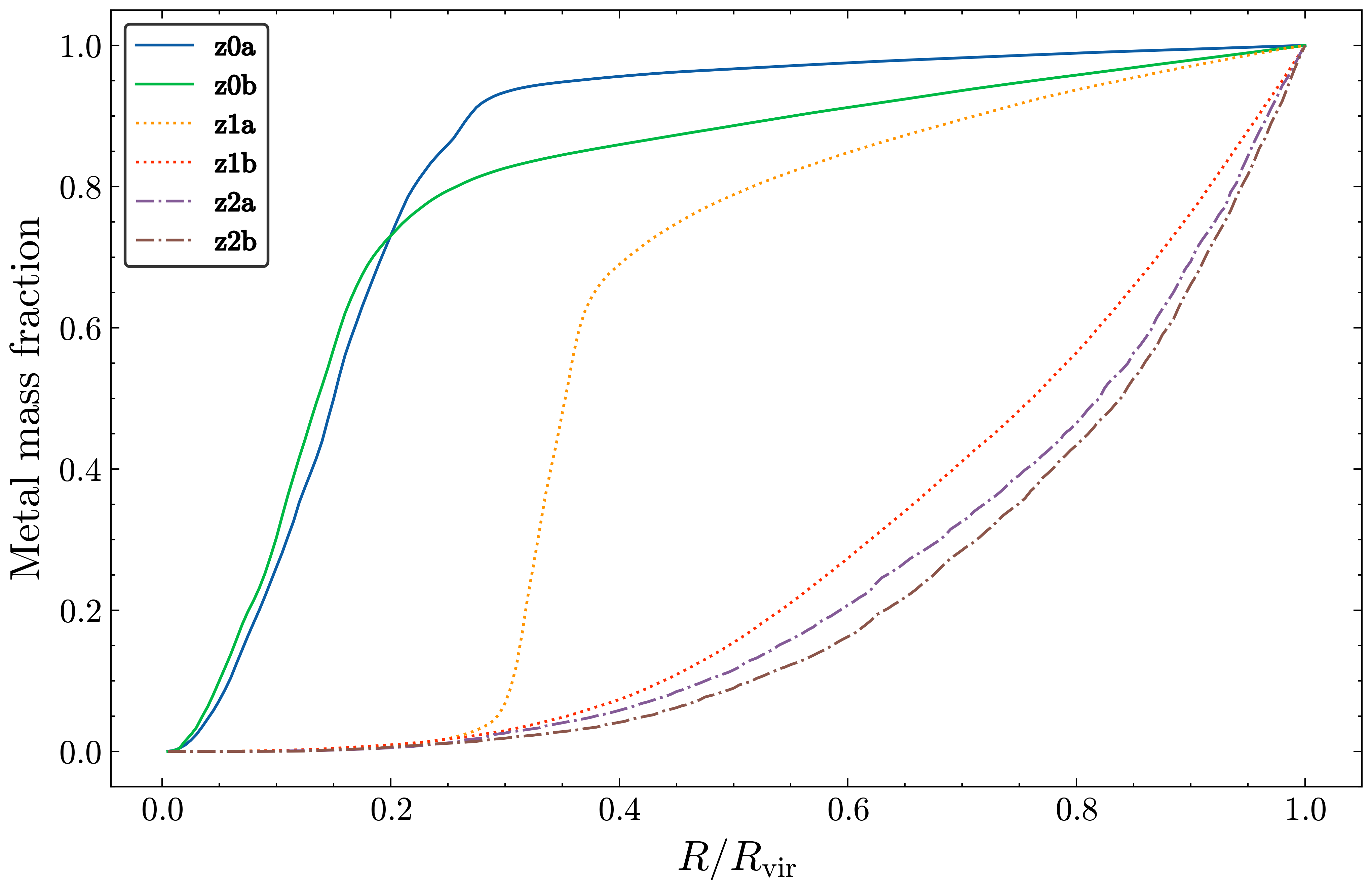} 
    \caption{Profiles of an accumulated metal fraction within $R_\mathrm{vir}$. For \textbf{z0a} and \textbf{z0b}, most of the metal resides in the galaxy, and their profiles start to diverge at $0.2 - 0.3 \, R_\mathrm{vir}$. Meanwhile, \textbf{z2a} and \textbf{z2b} show a different pattern, with most of the metal smoothly distributes around the CGM. The profiles of \textbf{z1a} and \textbf{z1b} fall between $z=1$ and $z=2$ models. The metal fraction in \textbf{z1b} is closer to the distribution observed in the $z = 2$ models, while the profile of \textbf{z1a} appears as a transition between \textbf{z0b} and \textbf{z1b}. Such transitions in the metal distribution may be related to the coevolutionary stage of the DGs and their environments.}
    \label{fig:metal_fraction}
\end{figure}

\cite{Bordoloi2014} suggested that the total carbon mass within a DG is comparable to that in its CGM at $z \sim 0$, which is consistent with the metallicity results of our simulations.   
We present the accumulated metal-mass fraction as a function of the radius within a halo in Figure \ref{fig:metal_fraction}. The profiles vary significantly between $z = 0$ and $z = 2$ halos.  
These differences relate to two different modes of metal enrichment.  
At $z = 0$, a steeper metallicity gradient between the galaxies and their CGM appears due to the weaker stellar and AGN feedback.
Furthermore, \textbf{z0a} and \textbf{z0b} can be explained by the "three-zone" structure, galaxies, inner CGM, and outer CGM, proposed in \citet{2021Li}, and they also agree with observational metal detections found within the half-virial radius for DGs at $z < 0.3$ \citep{2024Zheng}.
However, the feedback is much stronger at $z=2$ due to active star formation and SMBH accretion. It drives a strong outflow containing metal-rich gas from galaxies to CGM and smoothens the galaxy-CGM metallicity gradient. 
The flow driven by stellar and AGN feedback plays a crucial role in transporting the metal from galaxies to their CGM \citep{1986Dekel, 2022Sales}.
For high-redshift galaxies, stronger outflows can ship galactic metals to the CGM, even to the IGM. For the CGM region, the models of the larger virial mass models of \textbf{z0b}, \textbf{z1b}, and \textbf{z2b}, have a higher metal fraction than models of \textbf{z0a}, \textbf{z1a}, and \textbf{z2a} at a given $r$. We summarize the SFR and gas accretion rates contributed by the initial galaxy, CGM, and IGM in Table \ref{tb:acc_SFR}.
DGs of higher virial mass have higher SFR and gas accretion rates that apply to models of $z = 0$ and $z = 1$. 
Furthermore, in Figure \ref{fig:metal_fraction}, halos with a higher virial mass exhibit a higher metal fraction in CGM at $z<2$.  

In \cite{2024Zheng}, they found that only $\sim 10 \%$ of the metal in the whole halo exists in the warm phase (T $\sim 10^4$ K) of the CGM in the low-redshift DGs, which agrees well with our results of the warm phase metal with $6 \%$ for \textbf{z0a} and $15 \%$ for \textbf{z0b}. 
\cite{2024Zheng} also suggested that there is more metal in the warm phase as $z$ increases.
Meanwhile, our results show that the metal in the warm phase is $12 \%$(\textbf{z1a}) and $20\%$ (\textbf{z1b}) for $z = 1$, and $20 \%$(\textbf{z2a}) and $14 \%$(\textbf{z2a}) at $z = 2$.

% simulation
In our simulations, the mass of the CGMs over the baryonic mass of the halo is $16\%$(\textbf{z0a}), $38\%$(\textbf{z0b}) $35\%$(\textbf{z1a}) $42\%$(\textbf{z1b}) $50\%$(\textbf{z2a}) and $48\%$(\textbf{z2b}); it shows an increasing trend from $z = 0$ to $z = 2$ emphasizing the importance of CGM gas at higher $z$, as shown in Figure \ref{fig:mass_contribution}. 
Our results also align with the results from \cite{2019Hafen} based on FIRE collaborations \citep{Hopkins_2014, Hopkins_2018, 2023Hopkins}, suggesting that the CGM of $10^{10}-10^{12} \, \Ms$ halos contains $\sim40\text{–}70\%$ of its baryonic mass at $z = 2$.
The only outlier, \textbf{z0a}, shows a CGM mass fraction of $16\%$, a rare isolated environment where most baryons are concentrated in the galaxy.
In general, the mass fraction of the CGM gas in our models is slightly lower than the results of \cite{2019Hafen}.
This discrepancy may arise from differences between the simulation setup, e.g., the initial conditions of zoom-in runs and our semi-cosmological runs. Our TNG initial conditions contain stars formed before the simulations start, which may lead to higher stellar masses affecting the stellar feedback and gaseous halos.
\citet{2022Sales} also suggests that the difference in models may result from more baryons accumulating inside the galaxy in cosmological simulations. 
For gas mass alone, the CGM mass fractions over the total halo gas are $20\%$(\textbf{z0a}), $45\%$(\textbf{z0b}), $44\%$(\textbf{z1a}), $49\%$(\textbf{z1b}), $67\%$(\textbf{z2a}), $72\%$(\textbf{z2b}) in our simulation.

Based on FIRE and FIRE2 simulations, \citet{2019Hafen, 2020Hafen} suggests that $\sim60\text{--}80 \%$ of the CGM gas mass in $ 10^{10}\text{--}10^{12} \, M_\odot$ halos is fueled by IGM accretion, which is consistent with our findings of $60 \text{--} 70\%$ from \textbf{z2a} and \textbf{z2b}.

\subsection{Comparison with TNG50-1}

%We compare  as shown in Table \ref{tb:compare_tng}. 
The major difference between our models and their original counterparts in TNG50-1 is in their SF properties. Our galaxies exhibit higher SFRs within a smaller SF region, resulting in smaller $R_{\star,1/2}$. This discrepancy arises due to the combined effects of resolution, cooling, and the subgrid SF modeling. The original TNG projects are designed to study the galaxy formation at $z=0$ through cosmological simulations. Therefore, the ISM physics and SF processes are modeled with subgrid models with limited resolution by utilizing an effective equation of state in the ISM \citep{Springel2003, 2013Vogelsberger, 2014Torrey}. In contrast, our high-resolution simulations can resolve the clumpy SF regions and consider the detailed gas cooling and chemistry that can better model the SF processes. Furthermore, the enhanced feedback from the higher SFR can create compact galactic disks by evaporating the outer regions of less dense gas.
%fragmentation of smaller-scale gas clumps. 
%This leads to higher SFRs and more compact SF regions. In addition, higher resolution facilitates more efficient cooling, enhancing gas accretion and increasing the total gas mass within the galaxy ($M_{\mathrm{gas}}$). 

% maintaining gas temperatures above 10,000 K and 
The BH accretion rates in our models and TNG are also different. The gas accretion rates in our models are lower due to a weaker dynamical friction of the lower cell mass \citep{Lin2023}. As shown in Figure \ref{fig:smbh_acc}, the accretion history in our models shows episodic and bursty patterns due to the accreting of clumpy gas in the galactic centers. This accretion pattern agrees well with \citet{2013Gabor, 2017DeGraf}, demonstrating the effect of simulation resolution in modeling BH accretion.

%resolve small dense gas clumps and the resulting the BH as in Figure \ref{fig:smbh_acc}, triggering strong outflows that effectively remove the outer disk.
%rated are lower in our models due to fewer dynamical effects from large gas cells. Resolution differences also influence the  However, because 
%} 
% In this study, we excluded dwarf galaxies undergoing major mergers or ongoing interactions. The simulation box size is sufficiently large for zoom-in simulations, following the criteria outlined by Onorbe et al. (2013). We also carefully examined the merger histories to ensure that no major mergers occurred during the evolution of our target galaxies. Therefore, the impact of ongoing interactions should be minimal in our models.

\begin{deluxetable*}{l|ccccc}
\label{tb:compare_tng}
\tablecaption{Physical properties of DGs in our runs and TNG50-1 \label{tb:galaxies}}
\tablewidth{0pt}
\tablehead{
\colhead{Model} & \colhead{SFR} & \colhead{$M_\star$} & \colhead{$M_{\mathrm{gas}}$} & \colhead{$R_{\star,1/2}$} & \colhead{$M_\text{BH}$} \\
\colhead{} & \colhead{[$M_\odot \, \text{yr}^{-1}$]} & \colhead{[$10^8 \, M_\odot$]} & \colhead{[$10^8 \, M_\odot$]} & \colhead{[kpc]} & \colhead{[$10^6 \, M_\odot$]}
}
\startdata
\textbf{z1a}   & $0.36$  & $2.46$  & $3.15$  & $1.28$ & $1.75$ \\
\textbf{z1a*}  & $0.0078$ & $1.44$ & $0.44$  & $2.22$ & $2.45$ \\
\hline
\textbf{z1b}   & $0.18$  & $4.38$  & $0.016$ & $0.8$  & $1.60$ \\
\textbf{z1b*}  & $0.061$ & $1.62$  & $0.56$  & $1.44$ & $3.14$ \\
\hline
\textbf{z2a}   & $6.02$  & $19.6$  & $18.4$  & $1.54$ & $1.97$ \\
\textbf{z2a*}  & $0.69$  & $4.93$  & $4.33$  & $3.88$ & $3.87$ \\
\hline
\textbf{z2b}   & $4.35$  & $16.2$  & $6.59$  & $1.24$ & $1.86$ \\
\textbf{z2b*}  & $0.86$  & $6.85$  & $3.96$  & $3.37$ & $4.87$ \\
\enddata
\tablecomments{
We compare our simulated $z=1,2$ DGs with their original counterparts in TNG50-1 at the same epoch, which corresponds to the $z$ evolution from $\approx 2 \rightarrow  1.3$ and $\approx 1 \rightarrow  0.4$. 
 The DGs in TNG50-1 are marked with a star (*). Five physical quantities are listed: SFR, $M_\star$, $M_{\mathrm{gas}}$, $R_{\star,1/2}$, and $M_\text{BH}$.}
\end{deluxetable*}
% Comparison our simulated galaxies at $z = 1$ and $z = 2$ (models \textbf{z1a} to \textbf{z2b}) to the results from TNG50-1 at similar evolutionary times, approximately $z = 0.4$ and $z = 1.3$, respectively.

% \textbf{z1a}   & $0.36$  & $4.92$  & $3.15$  & $1.28$ & $1.75$ \\
% \textbf{z1a*}  & $0.0078$ & $2.88$ & $0.44$  & $2.22$ & $2.45$ \\
% \hline
% \textbf{z1b}   & $0.18$  & $8.75$  & $0.016$ & $0.8$  & $1.60$ \\
% \textbf{z1b*}  & $0.061$ & $3.24$  & $0.56$  & $1.44$ & $3.14$ \\
% \hline
% \textbf{z2a}   & $6.02$  & $39.1$  & $18.4$  & $1.54$ & $1.97$ \\
% \textbf{z2a*}  & $0.69$  & $9.86$  & $4.33$  & $3.88$ & $3.87$ \\
% \hline
% \textbf{z2b}   & $4.35$  & $32.4$  & $6.59$  & $1.24$ & $1.86$ \\
% \textbf{z2b*}  & $0.86$  & $13.7$  & $3.96$  & $3.37$ & $4.87$ \\

\subsection{limits of the Current Models}
\label{subsec:PossibleImprovement}
%We acknowledge the limitations of our simulations in this section. 
Although the resolution of the simulation is higher compared with previous galaxy simulations, we are still unable to resolve regions such as the star formation sites, stellar feedback, and the accretion disk of SMBH. Therefore, we employ subgrid models for star formation and SMBH accretion, with the convergence of these models carefully verified. Nevertheless, modeling the physics of stellar and SMBH feedback is still crude in the current simulations and has ample room to improve. Furthermore, we also adopt a local UV background for all of our simulations. This setup is a trade-off between maintaining the integrity of the local environment and accounting for the effect of UV photons originating from outside of our target DG. However, a more realistic UV background involving the local radiative field is needed to properly model the evolution of IGM and CGM.

\section{Conclusion}

We present new high-resolution simulations of DGs and their CGM/IGM across cosmic time with initial conditions from a realistic cosmological simulation of the \texttt{IllustrisTNG}. Our results show the complex multiphase gas of CGM and IGM driven by gas accretion, SMBH, and stellar feedback from the host galaxies. 
We also tracked the accretion gas from the CGM and IGM to quantitatively evaluate their impact on the evolution of galaxies. 
In general, the gas accreted from the CGM depends on $z$ and plays a crucial role in fueling the gas reservoir and driving galactic star formation. During 1.5 Gyr of evolution, the DGs grow solely through the accretion contributed by the CGM at $z = 0$ and $z = 1$. Overall, the CGM gas contributes $20 \% - 50 \%$ to the total star-forming gas and $40 \% - 70 \%$ gas mass of galaxies. At $z = 2$, CGM and IGM contain $60 \% - 70 \%$ of the total halo gas mass, and the consequent gas accretion from them further contributes $\sim 60 \%$ of the galaxy mass. The strong accretion flow from the IGM/CGM triggers intense AGN activity and active star formation at the galactic center. Furthermore, SMBH of the DGs at $z=2$ shows an episodic accretion history with a peak up to $\sim 10\%$ of the Eddington mass accretion rate indicating the phases of rapid growth. Our results suggest the increasing importance of the coevolution of DGs and their CGM in a high-$z$ universe possibly examined by the coming observations from the James Webb Space Telescope (JWST). 
%\citep{Zou2024}.

\acknowledgments
% \section*{Acknowledgments} 
The authors appreciate Chorng-Yuan Hwang for his insightful discussions and thank Chi-Hung Lin, Kung-Yi Su, and Po-Feng Wu for their support of this work. This research is supported by the National Science and Technology Council, Taiwan, under grant No. MOST 110-2112-M-001-068-MY3, NSTC 113-2112-M-001-028-, and the Academia Sinica, Taiwan, under a career development award under grant No. AS-CDA-111-M04. 
KC acknowledges the support of the Alexander von Humboldt Foundation and Heidelberg Institute for Theoretical Studies. This research was supported in part by the NSF PHY-2309135 grant to the Kavli Institute for Theoretical Physics (KITP) and the NSF PHY-2210452 grant to the Aspen Center for Physics. Our computing resources were supported by the National Energy Research Scientific Computing Center (NERSC), a U.S. Department of Energy Office of Science User Facility operated under Contract No. DE-AC02-05CH11231 and the TIARA Cluster at the Academia Sinica Institute of Astronomy and Astrophysics (ASIAA).

%\facility{facility ID}
%\facilities{facility ID, facility ID, facility ID}
%\software{Numpy}

\bibliographystyle{yahapj}
\bibliography{refs}

\end{document}